\documentclass[12pt,a4paper,final]{iopart}
\usepackage[utf8]{inputenc}
\usepackage{iopams}
\expandafter\let\csname equation*\endcsname\relax
\expandafter\let\csname endequation*\endcsname\relax
\usepackage{graphicx,dcolumn,bm,amssymb,amsmath,amsfonts,xcolor,mathtools}
\usepackage{subfigure}
\usepackage{dsfont}  
\usepackage{bbold}   
\usepackage{dutchcal} 
\usepackage[sort&compress,numbers]{natbib}
\usepackage[
pdfstartview=FitBV,
bookmarks=true,
bookmarksopen=true,
colorlinks =true,
linkbordercolor=blue,
linkcolor = blue,
citecolor = blue,
urlcolor=blue]{hyperref}
\usepackage[capitalize]{cleveref} 
\usepackage{multirow}

\crefname{figure}{Figure}{Figures}

\begin{document}
	
	\title[]{Mapping atomic trapping in an optical superlattice \\
		onto the libration of a planar rotor in electric fields}
	
	\author{Marjan Mirahmadi$^1$, Bretislav Friedrich$^1$, Burkhard Schmidt$^2$ and Jes\'us P\'{e}rez-R\'{i}os$^{1,3,4}$}
	\address{$^1$ Fritz-Haber-Institut der Max-Planck-Gesellschaft, Faradayweg 4-6, D-14195 Berlin, Germany}
	\address{$^2$ Institute for Mathematics, Freie Universit\"{a}t Berlin, Arnimallee 6, D-14195 Berlin, Germany}
	\address{$^3$ Department of Physics and Astronomy, Stony Brook University, Stony Brook, NY 11794, USA}
	\address{$^4$ Institute for Advanced Computational Science, Stony Brook University, Stony Brook, NY 11794, USA}
	\ead{mirahmadi@fhi-berlin.mpg.de, burkhard.schmidt@fu-berlin.de, bretislav.friedrich@fhi-berlin.mpg.de, jesus.perezrios@stonybrook.edu}
	
	\begin{abstract}
	We show that two seemingly unrelated problems -- the trapping of an atom in an optical superlattice (OSL) and the libration of a planar rigid rotor in combined electric and optical fields -- have isomorphic Hamiltonians. Formed by the interference of optical lattices whose spatial periods differ by a factor of two, OSL gives rise to a periodic potential that acts on atomic translation via the AC Stark effect. The latter system, also known as the generalized planar pendulum (GPP), is realized by subjecting a planar rigid rotor to combined orienting and aligning interactions due to the coupling of the rotor's permanent and induced electric dipole moments with the combined fields. The mapping makes it possible to establish correspondence between concepts developed for the two eigenproblems individually, such as localization on the one hand and orientation/alignment on the other. Moreover, since the GPP problem is conditionally quasi-exactly solvable (C-QES), so is atomic trapping in an OSL. We make use of both the correspondence and the quasi-exact solvability to treat ultracold atoms in an optical superlattice as a semifinite-gap system. The band structure of this system follows from the eigenenergies and their genuine and avoided crossings obtained previously for GPP as analytic solutions of the Whittaker-Hill equation. These solutions characterize both the squeezing and the tunneling of atoms trapped in an optical superlattice and pave the way to unraveling their dynamics in analytic form.
	\end{abstract}

	\section{Introduction}\label{sec:intro}	
	
	The spatial patterns imprinted upon ensembles of gaseous atoms by optical lattices have served as platforms for quantum simulation of condensed matter systems as well as for quantum information processing, including quantum computation \cite{Deutsch2002,Kay2004,Oberthaler2006,Lee2007,Folling2007,Esslinger2010,Wirth2011,Dutta2015,Li2016,Kangara2018,Schafer2020,Stamper_Kurn2020,Semeghini2021}. Superimposed commensurate lattices (or superlattices for short) whose spatial periods are in integer ratios have enabled patterned loading  key to achieving versatile atom-lattice architectures \cite{Peil2003}, quantum computing with atom transport \cite{Calarco2004}, atom-pair manipulation \cite{Sebby2006},  and topologically protected transport \cite{Lohse2016}. The engineering of optical lattices and superlattices has been recently reviewed in Ref. \cite{Windpassinger2013}. 
	
	Herein, we show that the translational confinement of atoms in an optical superlattice (OSL) formed by the interference of optical lattices whose spatial periods differ by a factor of two can be mapped onto the libration of a generalized planar pendulum (GPP). GPP is realized by subjecting a planar rotor to combined orienting and aligning interactions that arise due to the coupling of the rotor's permanent and induced dipole moments with collinear external electric fields \cite{Friedrich1991,Friedrich1991a,Seideman1995,Friedrich1999,Henriksen1999,Cai2001,Stapelfeldt2003,Seideman2005,Daems2005,Kiljunen2005,Leibscher2009,Rouzee2009,Schmidt2014a,Koch2019}. 
	
	\begin{figure}[b]
		\centering
		\includegraphics[scale=0.4]{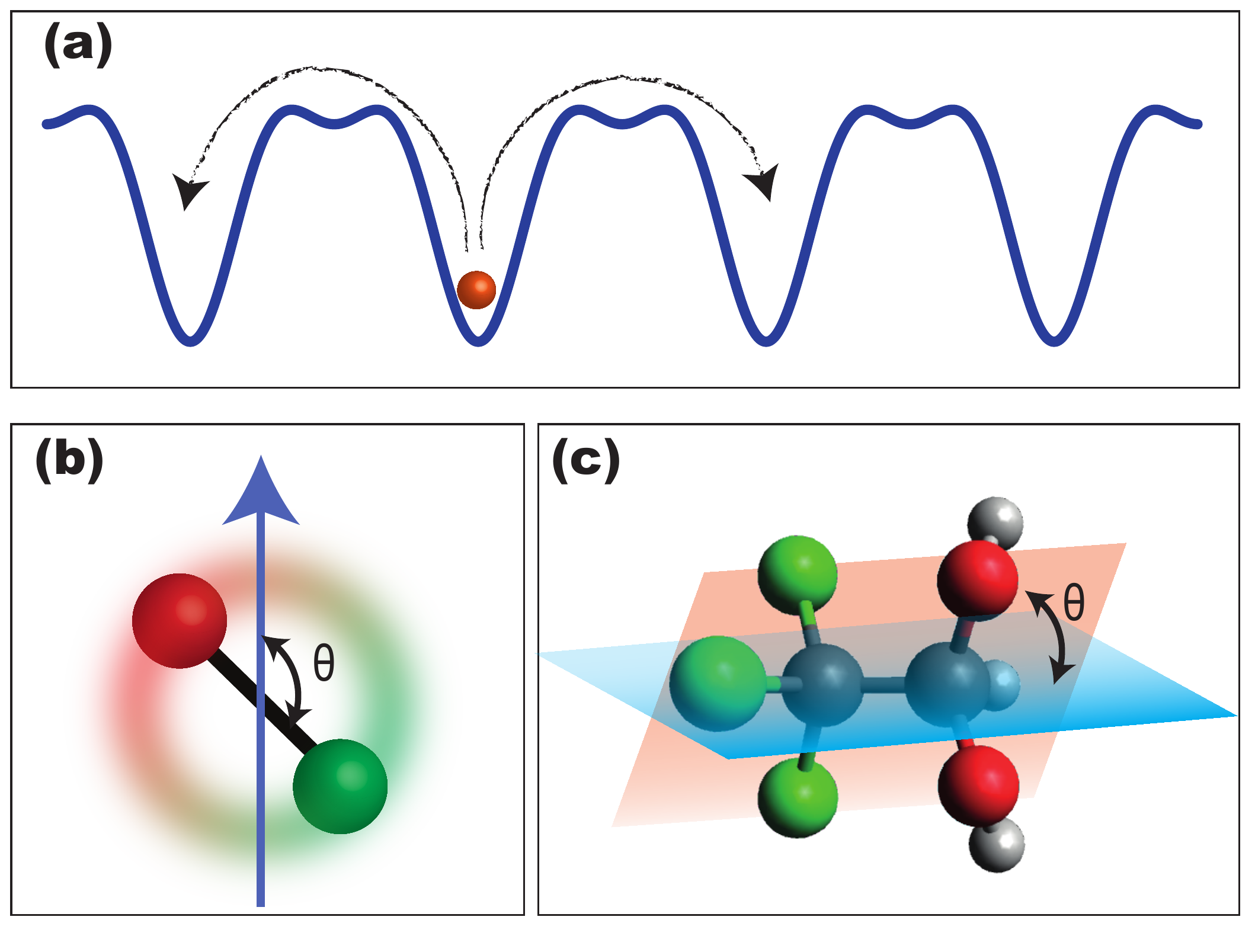} 
		\caption{\label{fig:intro} Three different eigenproblems whose Hamiltonians are isomorphic:  (a) trapping of ultracold atoms in an optical superlattice; (b) libration of a generalized planar pendulum in collinear external electric fields; (c) the torsional motion in a molecule such as CCl$_3$CH(OH)$_2$.} 
	\end{figure} 

	Interestingly, pulsed optical traps had been used earlier to simulate the kicked rotor \cite{Moore1995,Ammann1998,Klappauf1998} as a way of modelling quantum chaos and Anderson localization \cite{Fishman1982,Grempel1984,Izrailev1990,Tian2010,Hamilton2022}. In contradistinction, our present study makes use of the previously established features of the generalized quantum pendulum to shed light on the behavior of ultracold atoms confined in an optical superlattice. Among these features is the conditional quasi-exact solvability (C-QES) of the GPP eigenproblem: under certain conditions (i.e., for particular ratios of the strengths of the orienting and aligning interactions), many (but not all) eigenfunctions and eigenenergies are available in analytic form. These can be then used, \emph{inter alia}, to obtain the band structure of the atoms in the OSL or to establish relationships between the main physical characteristics of the two eigenproblems such as localization on the one hand and directionality (orientation and alignment) on the other.
	
	We note that also the Hamiltonian for molecular torsion in polyatomics \cite{Herschbach1959}, whether or not subject to coherent control \cite{Roncaratti2010,Parker2011,Ashwell2015}, is isomorphic with the Hamiltonians of the OSL and GPP systems, see \cref{fig:intro}. However, in what follows we focus on the OSL and GPP systems only.

	This paper is organized as follows: In \cref{sec:superlattice}, we introduce the Hamiltonian of a single atom subject to an optical superlattice. The isomorphism of this Hamiltonian with that of the generalized planar pendulum is established in \cref{sec:analogy}. In \cref{sec:CQES}, we provide a survey of the conditional quasi-exact solvability of the Schr\"{o}dinger equation for either Hamiltonian. In \cref{sec:semigap}, we make use of the spectral properties of the GPP system to investigate the band structure of the atoms trapped in an optical superlattice. The spatial localization of the band-edge Bloch states and its relation to the orientation and alignment of the pendulum is treated in \cref{sec:Sloc}. \cref{sec:delta_phi} evaluates the effect on the atomic confinement of the relative phase of the two optical lattices that make up the superlattice.   Finally, \cref{sec:concl} provides a summary of the present work and outlines prospects for its future applications. \ref{App1} details the analytically obtainable band-edge states while \ref{App2} outlines the spectral properties of the Hill equation.
	
	\section{An atom interacting with a one-dimensional optical superlattice}\label{sec:superlattice}
	
	A one-dimensional (1D) optical lattice, generated by the interference of two linearly polarized laser beams of the same wavelength $\lambda$ counter-propagating along the $x$ axis, produces, via the AC Stark effect, an optical trapping potential for atoms that is proportional to $\cos^2(kx)$, with $k=2\pi/\lambda$ the wave-number of either of the laser beams. Superimposing two such optical lattices, characterized by wavevectors $k_i=2\pi/\lambda_i$ with $i = 1,2$, leads to a superlattice that produces an optical potential \cite{Peil2003,Kay2004, Calarco2004,Kangara2018}
	\begin{align}\label{eq:supL}
		V(x) =V_0+ V_1 \cos(2k_1x) + V_2\cos(2k_2x-\varphi) 
	\end{align}
	with $V_i= d^2 \mathcal{E}_i^2/(2\hbar \Delta_i) $ the depth of the 1D lattice $i$, $\mathcal{E}_i$ the amplitude of the corresponding electric vector of the laser field, $d$ the projection of the atomic dipole moment $\vec{d}$ on the electric field $\vec{\mathcal{E}}_i$ (note that $d$ can be different for each lattice based on the atomic states involved), $\Delta_i $ the detuning of the laser field $i$ from the nearest atomic resonance, and $\hbar$ the reduced Planck constant. The relative phase of the two superimposed lattices is characterized by the angle $\varphi$. The constant AC Stark shift $V_0$ between the two constituent lattices $i=1$ and $i=2$ will be omitted. 
	
	Provided the laser fields are sufficiently far detuned from any atomic resonance, i.e., $\Delta_i \gg \Gamma$, with $\Gamma$ the spontaneous emission rate, we can invoke the adiabatic approximation \cite{Steck2019} and write the effective Hamiltonian for atoms in a 1D superlattice as\footnote{In this scenario, the atom can be treated as a two-level system whose evolution is described by that of its ground state.} 
	\begin{align}\label{eq:Heff}
		H_\mathrm{OSL} = -\frac{\hbar^2}{2m}\frac{d^2}{dx^2}  + V(x) 
	\end{align}
	with $m$ the atomic mass. Furthermore, as long as the scattering length of the atoms is small compared to $\lambda_i$, we can treat the atoms as one.  
	
	In what follows, we consider a superlattice generated by the interference of two optical lattices whose spatial periods differ by a factor of two, i.e., $ k_s \equiv k_1 = 2k_2$, where the subscript $s$ labels the lattice with the shorter wavelength, $\lambda_s$. For now, we set the relative phase $\varphi = 0$. However, in \cref{sec:delta_phi}, we consider the effect of a non-zero relative phase on the properties of the superlattice. 	
	Thus, \cref{eq:supL} can be recast in the form
	\begin{align}\label{eq:pot}
		V(x) = V_s \cos^2(k_sx) + V_{\ell}\cos(k_sx)
	\end{align} 
	which is suitable for establishing the mapping of the OSL onto the planar rigid rotor under the orienting ($\propto\cos$) and aligning ($ \propto\cos^2$) interactions, i.e., onto the GPP system. Note that in \cref{eq:pot}, we have neglected a constant shift of $ V_s/2$ due to the transformation from $\cos$ to $\cos^2$. The amplitudes $V_s$ ad $V_\ell$, with the subscript $\ell$ pertaining to the lattice with the longer wavelength, $\lambda_{\ell}$, are proportional to the depths of the ``short'' lattice, $V_1$, and ``long'' lattice, $V_2$, via 
	\begin{equation}\label{eq:vs}
		V_s= 2V_1  = \frac{\hbar\Omega_1^2}{4\Delta_1} 
	\end{equation}
	and 
	\begin{equation}\label{eq:vl}
		V_\ell= V_2 = \frac{\hbar\Omega_2^2}{8\Delta_2} 
	\end{equation}
	respectively, wherein $\Omega_i=-2\vec{d}\cdot\vec{\mathcal{E}_i}/\hbar$ is the Rabi frequency.  
	
\begin{figure}
		\centering
		\includegraphics[scale=0.45]{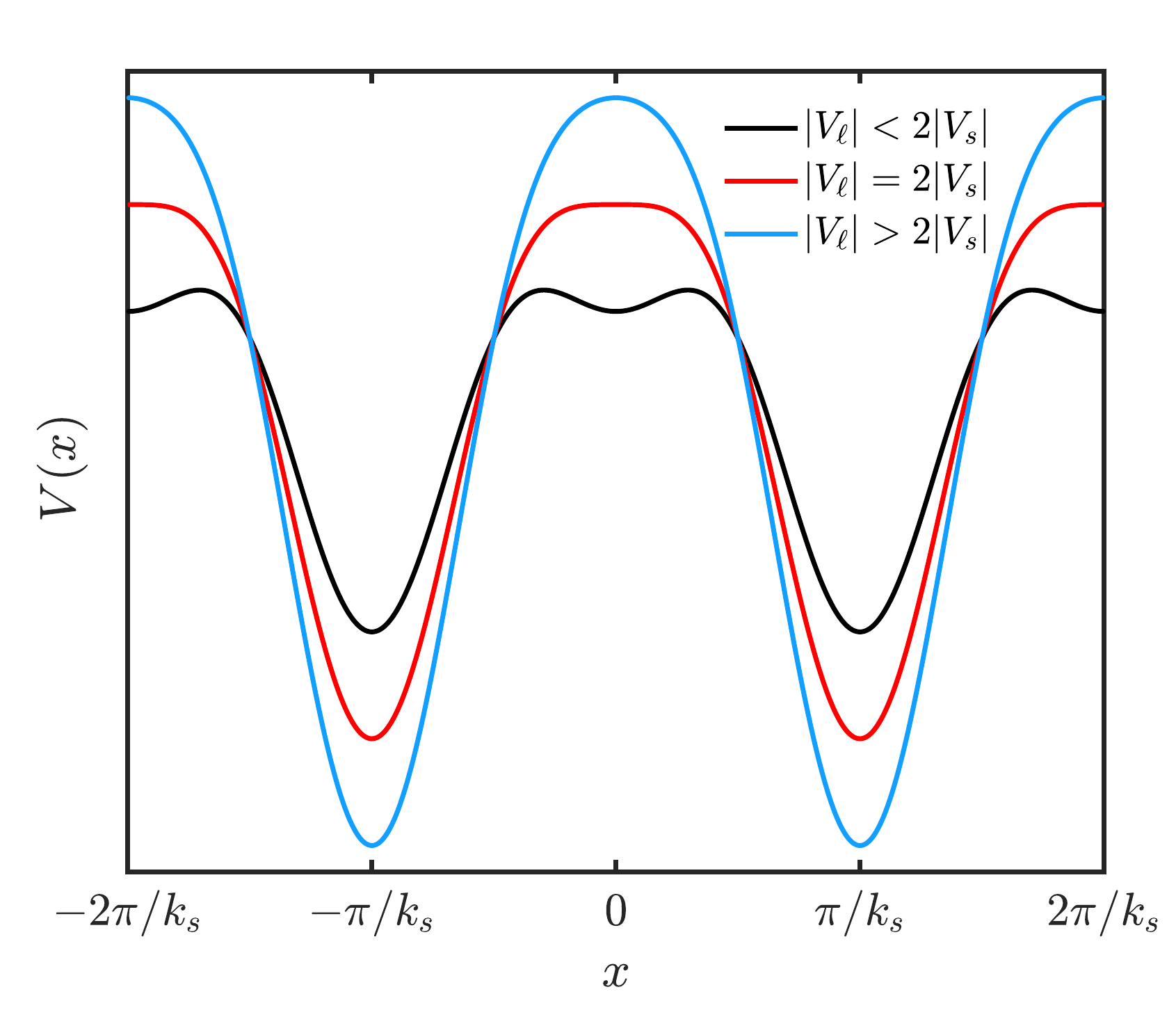} 
		\caption{\label{fig:pot1} Optical superlattice potential~\eqref{eq:pot} for different relative magnitudes of the parameters $V_\ell>0$ and $V_s<0$. Note that choosing $V_\ell<0$ results in a shift in $x$ of $V(x)$ by $\pi/k_s$.} 
	\end{figure}

\begin{figure}
		\centering
		\includegraphics[scale=0.6]{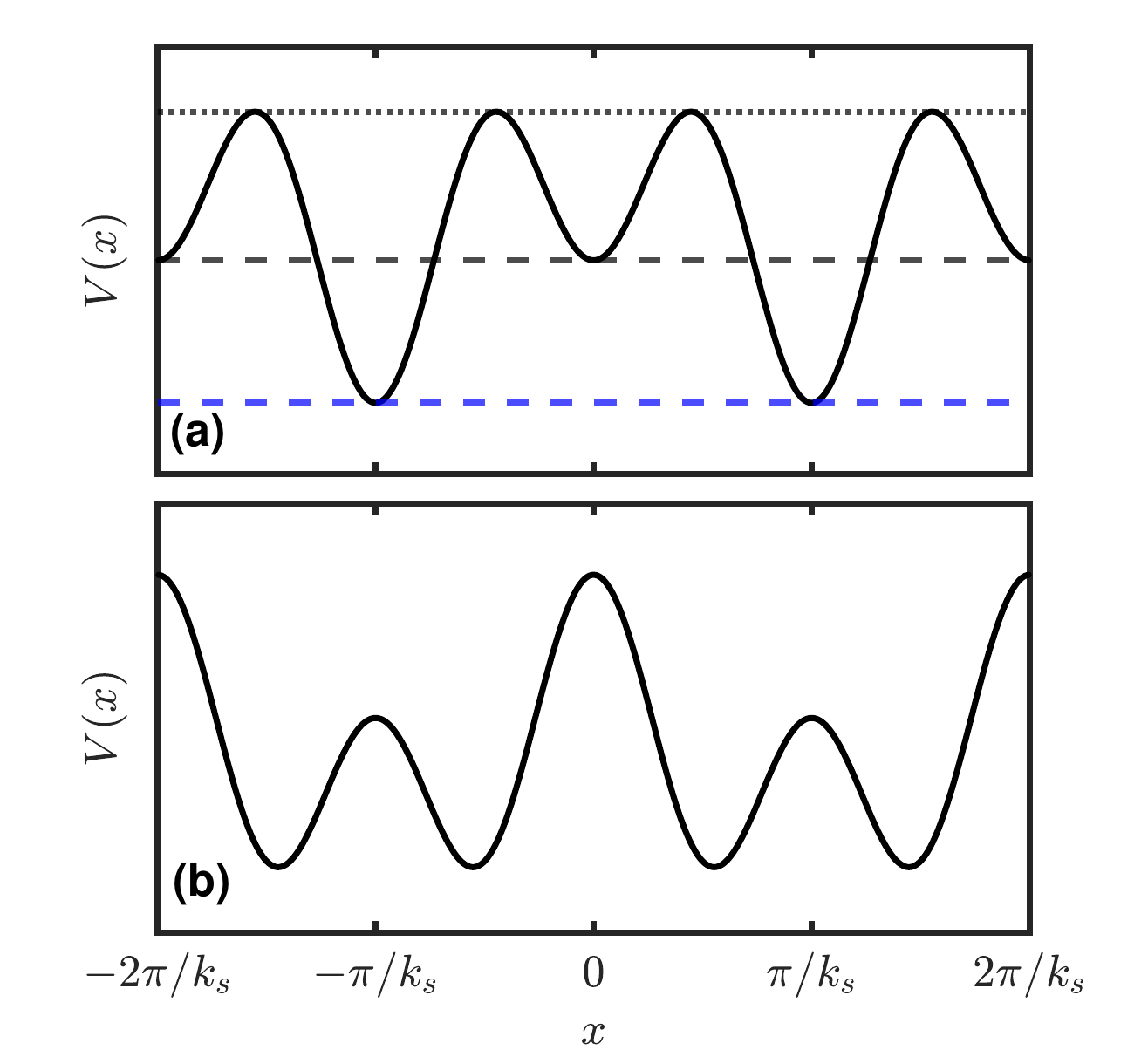} 
		\caption{\label{fig:pot2}Superlattice optical potential~\eqref{eq:pot} for $V_\ell> 0$ and (a) $V_s<0$ whose local minimum, global minimum, and maximum are marked by the black dashed, blue dashed and black dotted lines, respectively. (b) $V_s>0$. In either panel, $\left|V_\ell\right| < 2\left|V_s\right|$.} 
	\end{figure}

	The optical potential~\eqref{eq:pot} due to the superlattice is a periodic function with period (or ``lattice constant'') $a = 2\pi/k_s$ whose shape depends on the relative magnitude and sign of the amplitudes $V_\ell$ and $V_s$. As shown in \cref{fig:pot1}, for $V_s<0$ the shape of the OSL potential can be varied from a single-well (SW) potential in the case $|V_\ell|>2|V_s|$ to an asymmetric double-well (DW) potential when $|V_\ell|<2|V_s|$ over the unit cell of the superlattice. For $|V_\ell|=2|V_s|$, the potential has a flat maximum where the first, second, and third derivatives of the potential are zero. Hence the shape of the OSL potential for a given atom can be tailored by changing the ratio of the laser intensity to the detuning of the two constituent optical lattices.  
	The choice of the sign of $V_\ell$ is arbitrary since it is equivalent to a shift of $V(x)$ by half a period ($=\pi/k_s$) in $x$. Thus,  without a loss of generality, we can assume $V_\ell>0$, although the results and discussion presented below apply to both cases: for a blue-detuned ($\Delta_2>0$) as well as a red-detuned long lattice ($\Delta_2<0$). 	In contrast, the lattice geometry and its band structure are qualitatively different depending on whether $V_s$ is positive or negative, as illustrated in \cref{fig:pot2}. Hereafter, we consider the short lattice to have a red detuning, $\Delta_1<0$ ($V_s$ negative), giving rise to an OSL potential consisting of an asymmetric double well with a local minimum, $(V_s+V_\ell) $, a global minimum, $(V_s-V_\ell)$, and a maximum, $-V_\ell^2/(4V_s)$, as shown in panel~(a) of \cref{fig:pot2}. Panel (b) shows what the OSL potential looks like for $V_s$ positive.	
	
	\section{Comparing an atom subject to an optical superlattice with a planar rotor subject to combined orienting and aligning interactions a.k.a. planar generalized pendulum}\label{sec:analogy}
	
	The Hamiltonian of a planar (2D) rigid rotor subject to collinear orienting and aligning interactions is given by \cite{Schmidt2014,Becker2017,Mirahmadi2020} 
	\begin{equation}\label{eq:RotorH}
		H_\mathrm{GPP} = -B\frac{d^2}{d\theta^2} -\eta\cos\theta-\zeta\cos^2\theta 
	\end{equation} 
	where $B=\hbar^2/(2I)$ is the rotational constant, with $I$ the moment of inertia, and $0\le\theta <2\pi$ is the polar angle between the axis of the rotor and the direction of the external collinear fields. It is the couplings of the permanent and induced dipole moments, fixed to the axis of the rotor, with the external collinear fields that give rise to the orienting and aligning interactions, see panel (b) of \cref{fig:intro}. The strengths of the orienting and aligning interactions are characterized, respectively, by the parameters $\eta$ and $\zeta$. For either a vanishing $\eta$ or $\zeta$, the time-independent Schr\"odinger equation (TISE) pertaining to the resulting Hamiltonian becomes isomorphic with the Mathieu equation, which satisfies different boundary conditions for the purely orienting and purely aligning interactions, cf. Table III of Ref. \cite{Friedrich1991a}. When both $\eta$ and $\zeta$ vanish, the eigenproblem becomes that of a planar rotor. 
	
	
	
	In order to establish the isomorphism of Hamiltonians (\ref{eq:Heff}) and (\ref{eq:RotorH}), we introduce a dimensionless variable $\theta_a\equiv k_sx$ ($=2\pi x/a$) whose substitution transforms Hamiltonian (\ref{eq:RotorH}) into	
	\begin{align}\label{eq:H_sup}
		H_\mathrm{OSL} = -E_R\dfrac{d^2}{d \theta_a^2} + V_s \cos^2\theta_a + V_{\ell}\cos\theta_a
	\end{align}
	with $E_R = \hbar^2k_s^2/(2m)$ the atomic recoil energy. Note that the recoil energy is related to the lattice constant $a$ via $E_R = 2\pi^2\hbar^2/(ma^2)$. Comparing Hamiltonians \ref{eq:RotorH} and \ref{eq:H_sup} makes it possible to establish a correspondence between the interaction parameters of the OSL and GPP eigenproblems, see \cref{tab1}.
	\begin{table}
	\caption{\label{tab1} Correspondence between the interaction parameters of the OSL and GPP eigenproblems.}
		\begin{center}
			\begin{tabular}{@{}ccccc}
				\br
				System &  \multicolumn{4}{c}{Parameters} \\
				\br
				Atom in an optical superlattice (OSL) & $\theta_a$  &$V_s$ & $V_\ell$ &  $E_R$ \\ 
				Generalized planar pendulum  (GPP) & $\theta$  & $-\zeta$ & $-\eta$  & $B$ \\
				\br
			\end{tabular}
		\end{center}
	\end{table}
	
	Note that, unlike the polar angle $\theta$, the variable $\theta_a$ in \cref{eq:H_sup} is not defined on a circumference but on a line\footnote{In classical mechanics, the dynamics of a particle subject to a linear periodic potential is identical to that of a rotor, which is not the case for its quantum mechanical counterpart.}. In particular, the  interval $0\le\theta_a<2\pi$ describes a unit cell consisting of an asymmetric double-well with a local minimum at $\theta_a = 0$ or $2\pi$ and a global minimum at $\theta_a = \pi$. 	
	
	
	In order to explore the spectral properties of the OSL and GPP Hamiltonians (\ref{eq:Heff}) and (\ref{eq:RotorH}), we divide the Schr\"odinger equation pertaining to each Hamiltonian through its characteristic energy, $E_R$ or $B$. Thus, for the OSL system we obtain
	\begin{align}\label{eq:OPTISE} 
		\tilde{H}_\mathrm{OSL}\psi(\theta_a)&=E\psi(\theta_a)  
	\end{align}
	with $H_\mathrm{OSL}/E_R \rightarrow \tilde{H}_\mathrm{OSL}$. Hence the eigenvalues $E$ of $\tilde{H}_\mathrm{OSL}$ pertatining to eigenfunctions $\psi(\theta_a)$ are rendered in units of recoil energy $E_R $.
	
	On the other hand,  the reduced eigenvalue problem for the GPP system becomes	
	\begin{align}\label{eq:RotTISE} 
		\tilde{H}_\mathrm{GPP}\phi(\theta)&=\epsilon\phi(\theta)
	\end{align}
	with $H_\mathrm{GPP}/B \rightarrow \tilde{H}_\mathrm{GPP}$. The
	eigenvalues $\epsilon$ of $\tilde{H}_\mathrm{GPP}$ then come out in units of the rotational constant $B$. The corresponding eigenfunctions are $\phi(\theta)$.
	
	Despite the above similarity of the OSL and GPP eigenproblems, the physically meaningful boundary conditions on the two systems lead to different structures of the energy levels of \cref{eq:RotTISE,eq:OPTISE}. 
	
	Due to its spatial periodicity, \cref{eq:OPTISE} can be treated via Floquet's theorem (or equivalently Bloch's theorem), with solutions obeying the boundary condition
	\begin{align}\label{eq:blochBC}
		\psi(\theta_a+2\pi) = \mu\psi(\theta_a) 
	\end{align}
	where $\mu = \exp(i 2\pi q)$ is the Floquet multiplier and $2\pi\hbar q/a$ is the quasi-momentum. Consequently, the eigenvalues of \cref{eq:OPTISE} are energy bands $E \equiv E_{n}(q)$ with $n=0,1,2,\cdots$ the band index. Note that the parameter $q$ is continuous and confined to the first Brillouin zone (in the reduced-zone scheme), i.e., $-1/2\le q<1/2$. For physically meaningful solutions, the modulus of $\mu$ must be equal to one, i.e., the parameter $q$ must be real (see \ref{App2}). 
	
	In the case of the generalized planar pendulum, the TISE~\eqref{eq:RotTISE} may be solved either for a periodic boundary condition, 
	\begin{align}\label{eq:periodBC}
		\phi(\theta+2\pi) = \phi(\theta)
	\end{align}
	or an antiperiodic boundary condition\footnote{We include these $2\pi$-antiperiodic (or, equivalently, $4\pi$-periodic solutions as they may prove useful for problems involving Berry’s geometric phase or systems with $4\pi$ rotational symmetry.}, 
	\begin{align}\label{eq:aperiodBC}
		\phi(\theta+2\pi) = -\phi(\theta)
	\end{align}
	Given that \cref{eq:periodBC,eq:aperiodBC} are equivalent to \cref{eq:blochBC} for $\mu = 1$ and $\mu=-1$, respectively, the planar pendulum eigenstates correspond to Bloch waves for atoms in an optical superlattice with integer and half-integer wave numbers. In other words, the eigenfunctions and eigenvalues of the GPP Hamiltonian are equivalent to those at the edges of the first Brillouin zone: the periodic solutions to $q=0$ and the antiperiodic ones to $|q| = 1/2$.  	 
	
	\section{Conditional quasi-exact solvability (C-QES) of the time-independent Schr\"odinger equation (TISE)}\label{sec:CQES}
	
	The solvability of the TISE~\eqref{eq:RotTISE} as well as its spectral properties have been studied by means of supersymmetry and Lie-algebraic methods in our previous work \cite{Schmidt2014,Schmidt2014a,Becker2017,Schatz2018,Mirahmadi2020}. In this section, we make use of the results obtained therein to study the trapping of atoms in an optical superlattice. Based on the relation between Hamiltonians \cref{eq:RotTISE,eq:OPTISE}, we provide analytic insights into the band-gap structure of the optical superlattice. 
	
	
	The TISE~\eqref{eq:RotTISE} for the GPP system can be mapped onto the Whittaker-Hill differential equation \cite{liapounoff1902,Whittaker1914} (a special case of the Hill differential equation \cite{Magnus2004}),
	\begin{align}\label{eq:WHeq}
		\frac{d^2f(y)}{d y^2} + \left[\lambda + 4\kappa\beta\cos(2y) + 2\beta^2\cos(4y)\right]f(y) = 0
	\end{align}
	by making use of the definitions of the angular variable $y \equiv\theta/2$, the eigenvalues $\lambda \equiv (4\epsilon+2\zeta)$, and the real parameters $\kappa$ and $\beta$ via
	\begin{align}\label{eq:change_var}
		\eta/B  =  \kappa \beta ~ \quad  \zeta/B = \beta^2
	\end{align}
	In the same way, we can map \cref{eq:OPTISE} for the OSL system onto the Whittaker-Hill differential equation by setting $y \equiv\theta_a/2$, $\lambda \equiv (4E-2V_s)$, and
	\begin{align}\label{eq:change_varOP}
		V_\ell/E_R  =  -\kappa \beta ~ \quad  V_s/E_R = -\beta^2
	\end{align}
	The parameter $\kappa$ has been termed the \emph{topological index} \cite{Schmidt2014,Schmidt2014a}.     
	
	For $\zeta>0$, the GPP system belongs to the class of conditionally quasi-exactly solvable (C-QES) eigenproblems. This means that it is possible to obtain a finite number of its eigenvalues and eigenfunctions analytically (quasi-exact solvability, QES) \cite{Turbiner1988,Turbiner1988a,Shifman1989,Gonzalez-Lopez1996}), but only if the interaction parameters $\eta$ and $\zeta$ satisfy a particular condition (conditional exact solvability, CES) \cite{Dutt1995,Junker1998,Roychoudhury2001}). 
	Specifically, analytic solutions of \cref{eq:RotTISE} for the GPP system only obtain for integer values of the topological index $\kappa$. In addition, the integer values of $\kappa$ specify the number of obtainable analytic solutions. 
	
	Due to the GPP $\mapsto$ OSL mapping, we see that the band-edge wavefunctions and energies of the optical superlattice with TISE~\eqref{eq:OPTISE} are analytically obtainable only for integer ratios 
	\begin{equation}\label{eq:kappa}
		\kappa = \frac{|\eta|}{\sqrt{\zeta}} = \frac{|V_\ell|}{\sqrt{-V_s}} 
	\end{equation}
	Note that this statement is only valid for $V_s<0$ as TISE \eqref{eq:OPTISE} is not C-QES if the short-lattice is blue-detuned. 
	
	If $\kappa$ is an odd integer, the first $\kappa$ states obeying the periodic boundary condition (i.e., band-edge states with $q=0$ or integer wavenumbers) are analytically obtainable. If $\kappa$ is an even integer, the $\kappa$ lowest antiperiodic solutions (i.e., band-edge states with $|q|=1/2$ or half-integer wavenumbers) can be obtained analytically. In Refs. \cite{Becker2017,Mirahmadi2020}, analytic expressions for forty GPP eigenenergies $\epsilon(\beta)$ have been found. Those obtained for $\kappa = 1$ to $\kappa=4$ are listed in \cref{tab2} as band-edge energies $E(\beta)=\epsilon$. In addition, more details, including the analytic expressions for the band-edge eigenfunctions, are given in \ref{App1}.  
	\begin{table}
			\caption{\label{tab2}  Analytically obtained lowest band-edge energies of the optical superlattice (eigenenergies $\epsilon^{(\Gamma)}$) for the first four values of $\kappa$ defined in \cref{eq:kappa}. Here $\Gamma$ stands for the irreducible representations of the $C_{2v}$ point group of $H_{\mathrm{GPP}}$.}
		\centering
		\begin{tabular}{@{}ccc}
			\br
			$\kappa$           & $\Gamma$ &   $E=\epsilon^{(\Gamma)}$        \\
			\br
			1  & $A_1$ &  $-\beta^2$ \\  \mr
			\multirow{2}{*}{2} & $B_1$ &  $-\beta^2 - \beta + 1/4$ \\
			& $B_2$ & $-\beta^2 + \beta + 1/4$ \\ \mr
			\multirow{3}{*}{3} & $A_1$ & $-\beta^2 - \frac{1}{2}\sqrt{16\beta^2 + 1} + 1/2$  \\
			& $A_1$ & $-\beta^2 + \frac{1}{2}\sqrt{16\beta^2 + 1} + 1/2$  \\
			& $A_2$ &  $-\beta^2 + 1$ \\ \mr
			\multirow{4}{*}{4} & $B_1$ &  $-\beta^2 - \beta - \sqrt{4\beta^2 + 2\beta + 1} + 5/4$ \\
			& $B_1$ & $-\beta^2 - \beta + \sqrt{4\beta^2 + 2\beta + 1}  + 5/4$ \\
			& $B_2$ & $-\beta^2 + \beta - \sqrt{4\beta^2 - 2\beta + 1}  + 5/4$ \\
			& $B_2$ & $-\beta^2 + \beta + \sqrt{4\beta^2 - 2\beta + 1} + 5/4$ \\  
			\br		
				\end{tabular}
		\end{table}

	We note that if $\beta\rightarrow 0$ as $\kappa\rightarrow\infty$ and, at the same time, $\kappa\beta$ remains finite, then the Whittaker-Hill equation (\ref{eq:WHeq}) reduces to the Mathieu equation. As the above conditions of quasi-exact solvability do not apply to the Mathieu equation, it has no analytic solutions.   
	
	\section{Atoms subject to an optical superlattice as a semifinite-gap system}\label{sec:semigap}
	
	The spectrum of a periodic Schr\"odinger operator consists of regions of allowed eigenvalues (bands) where the corresponding eigenfunctions are bounded, and forbidden eigenvalues (gaps), where the eigenfunctions do not have a finite norm and, therefore, are not physically meaningful. As shown in \cref{sec:analogy}, for the TISE~\eqref{eq:OPTISE} of the OSL system, the bands only obtain for $q$ real (in which case $q$ corresponds to the Bloch wavenumber). The $q$ parameter as a function of $E$ can be determined from the Hill discriminant, $\mathcal{D}(E)$, by making use of the relation,
	\begin{equation}\label{eq:Dis_q}
		2\cos(2\pi q) =  \mathcal{D}(E)
	\end{equation}
	This procedure is commonly used to describe the band structure of a periodic differential equation such as the Hill equation \cite{Kohn1959,Kramers1935,Magnus2004,Teschl2012g}. In \ref{App2}, we summarize the procedure resulting in \cref{eq:Dis_q}, which is valid for any real and smooth periodic potential. 
	
\begin{figure}
		\centering
		\includegraphics[scale=0.65]{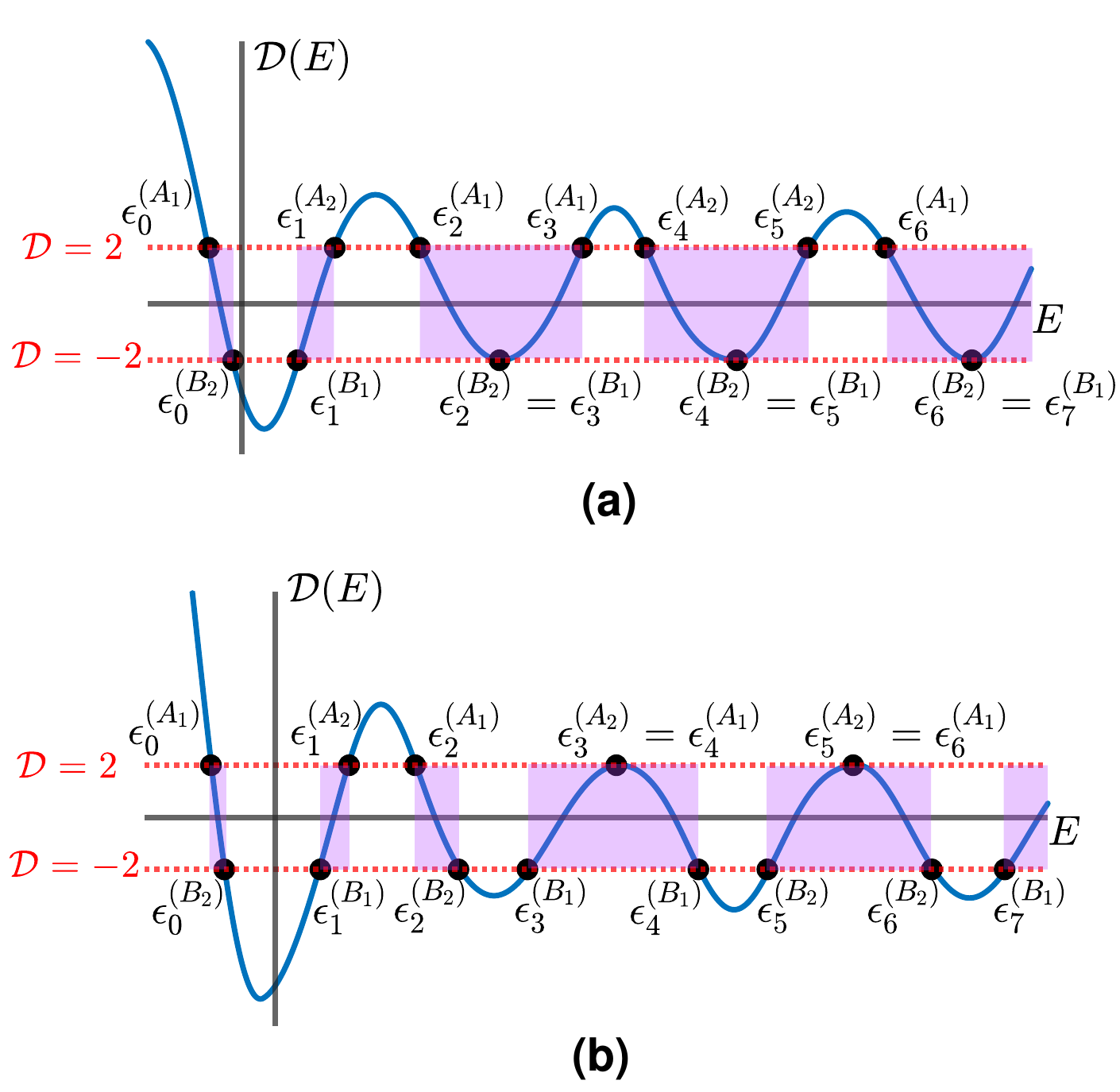} 
		\caption{\label{fig:HillD}Schematic diagram of the Hill discriminant for atoms in an optical superlattice treated as a semifinite-gap system for (a) $\kappa=2 $ and (b) $\kappa=3 $. The shaded areas indicate the allowed energy $(E)$ regions for $|\mathcal{D}(E)|\le2$. Note that the widths of the bands and gaps depend on the choice of the values of the $V_s$ and $V_\ell$ parameters.} 
	\end{figure}

	By making use of \cref{eq:Dis_q}, it is straightforward to locate the allowed and forbidden energy regions: if $q$ is real, $|\mathcal{D}(E)|\le2$, which defines the energy bands; if $q$ is not real, $|\mathcal{D}(E)|>2$, which defines the energy gaps. 
	In particular, the eigenvalues that satisfy $|\mathcal{D}(E)|=2$, define the band-edge states whose parameter $q$ takes integer or half-integer values. In general, the Hill discriminant is an oscillating function of the (real) variable $E$ that intersects the lines $\mathcal{D}(E)=\pm 2$ in the course of each oscillation. Consequently, the energy bands implied by the Hill equation obey the inequality $|\mathcal{D}(E)|\le2$.  The bands are separated by forbidden regions (gaps) where $|\mathcal{D}(E)|>2$. However, for the TISE~\eqref{eq:OPTISE} of the OSL system, after a few oscillations, the Hill discriminant intersects only one of these two $\pm2$ lines and, eventually, touches but one of them without crossing it, as depicted in \Cref{fig:HillD}. A system with a spectrum whose every second gap is eventually closed is referred to as a \emph{semifinite-gap system}~\cite{Correa2008,Hemery2010}.       
	While the optical superlattice with potential $V(x)$ of \cref{eq:pot} represents such a system, a system described by the Mathieu equation does not. 
	
	\begin{figure}
		\centering
		\includegraphics[scale=0.5]{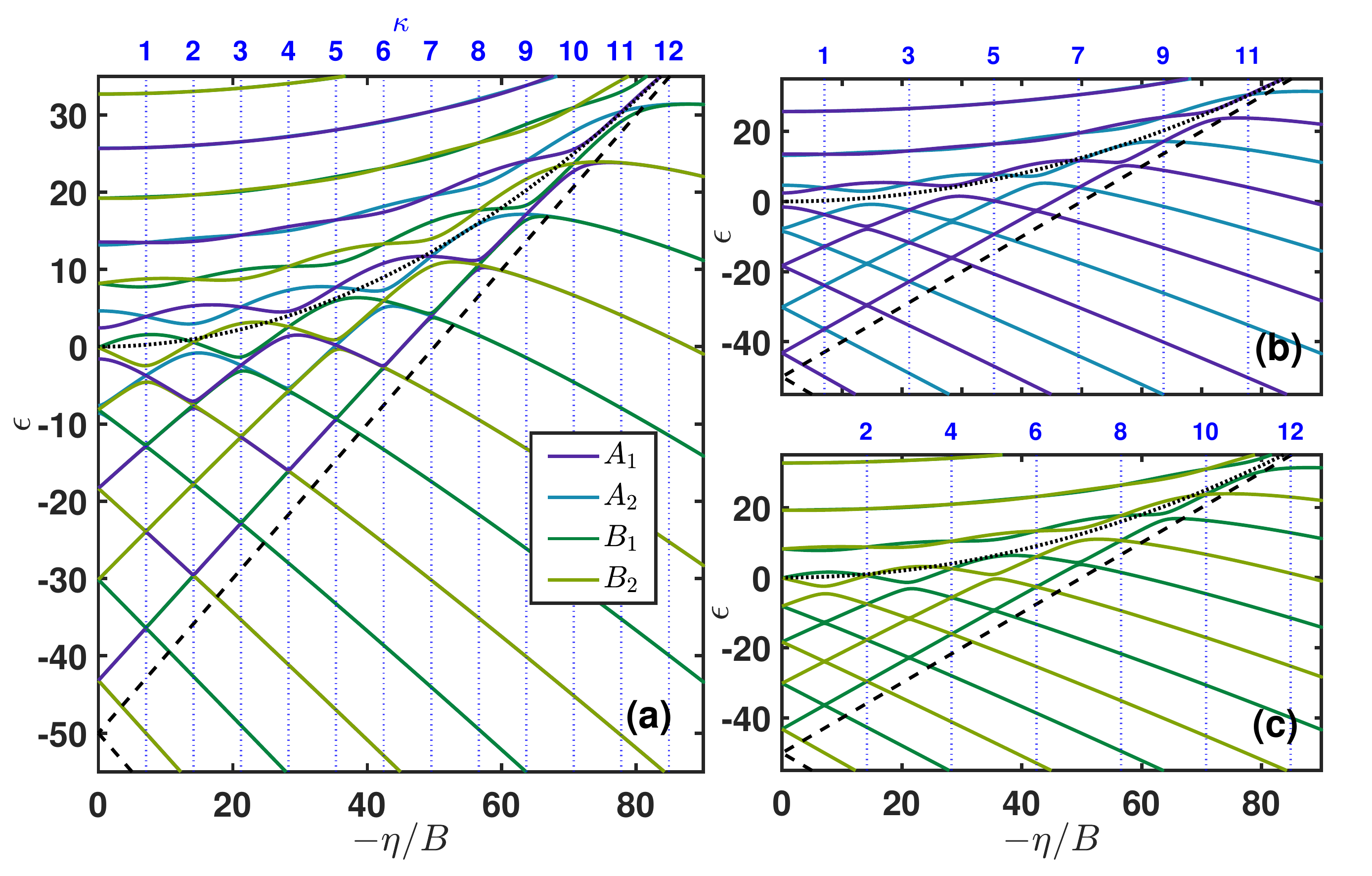} 
		\caption{\label{fig:zeta}(a) Energies of the generalized planar pendulum as functions of $\eta$ for constant $\zeta = 50~B$. The vertical blue dotted lines mark the $\eta$ values associated with the integer values of $\kappa$ from $\kappa=1$ $(\eta = -\sqrt{50})$ to 14 ($\eta = -14\sqrt{50}$). Panels (b) and (c) show the energies of the periodic ($A_1, A_2$) and antiperiodic ($B_1, B_2$) states, respectively. Here, the potential is an asymmetric double well with local and global minima indicated by black dashed lines, and a maximum shown by the black dotted curve.} 
	\end{figure} 
	
	Since the eigenvalues $E$ of the OSL system satisfying $\mathcal{D}(E)=\pm2$ correspond to the spectrum of GPP's TISE with periodic (+2) and antiperiodic (-2) boundary conditions, the knowledge of the GPP spectrum provides a new perspective on the band structure of ultracold atoms in an optical superlattice, as encapsulated in \cref{fig:HillD}. 
	
	As the symmetries of the generalized planar pendulum are isomorphic with those of the $C_{2v}$ point group \cite{Becker2017}, the solutions of the Schr\"odinger equation (\ref{eq:RotTISE}) for the GPP system fall into four categories, each corresponding to one of the irreducible representations $\Gamma\in~\{A_1,A_2,B_1,B_2\}$ of $C_{2v}$~\cite{Becker2017,Mirahmadi2020}. The solutions associated with the $A_1$ and $A_2$ symmetries are, respectively, even and odd functions (with respect to $\theta = \pi$), satisfying the periodic boundary condition on the interval $\theta \in[0,2\pi]$. The solutions corresponding to $B_1$ and $B_2$ symmetries are, respectively, odd and even functions (with respect to $\theta = \pi$), satisfying the antiperiodic boundary condition\footnote{Note that the correlation between the even functions and $B_2$ (or odd functions and $B_1$) is valid for $\eta<0$ and will change to the correlation between even functions and $B_1$ (odd functions and $B_2$) for $\eta>0$. For more details, see Refs.~\cite{Becker2017,Mirahmadi2020}}. In accordance with Sturm's oscillation theorem \cite{Magnus2004,Teschl2012g}, the eigenvalues form a monotonously increasing infinite sequence of real values $	\epsilon^{(A)}_0 < \epsilon^{(B)}_0 \leq \epsilon^{(B)}_1 < \epsilon^{(A)}_1 \leq \epsilon^{(A)}_2 < \epsilon^{(B)}_2 \leq \epsilon^{(B)}_3 < \ldots ~$, where $\{\epsilon^{(A)}_i\}$ is the energy set corresponding to the periodic solutions (either $A_1$ or $A_2$), and $\{\epsilon^{(B)}_i\}$ corresponds to the antiperiodic solutions (either $B_1$ or $B_2$). 
	Furthermore, the number of nodes of the corresponding eigenfunctions in the interval $[0,2\pi]$ is equal to $0,1,1,2,2,3,3,\ldots ~$, where the odd (even) number of nodes corresponds to antiperiodic (periodic) eigenfunctions.

	\Cref{fig:zeta} shows the energy levels of the generalized planar pendulum as a function of the orienting parameter $\eta$ for a constant value of the aligning parameter $\zeta = 50~B$. The energy levels that lie beyond the C-QES interval, i.e., above the local minimum of the potential (marked by the upper dashed lines), have been obtained numerically by means of the Fourier grid Hamiltonian method \cite{Marston1989} as implemented within the WavePacket software package~\cite{Schmidt2017,Schmidt2018,Schmidt2019}.  All eigenvalues below the local minimum of the potential, either analytic (integer $\kappa$) or numerical (non-integer $\kappa$), pertain to singlet states with a specific symmetry $\Gamma$. However, the energy differences between some pairs of the A and B levels in this part of the spectrum are small and hardly discernible on the scale of panel (a), which is why they are shown once more but separately: $A$ levels  in panel (b) and $B$ levels in panel (c).   
	Note that the ground state always pertains to the $A_1$ symmetry. 
	
	The most striking feature of the eigenenergies shown in \cref{fig:zeta} is their rich pattern of genuine and avoided crossings. As expected from the Wigner-von Neumann non-crossing rule \cite{VonNeumann1929,landau1991}, levels pertaining to the same symmetry (i.e., to the same irreducible representations $\Gamma$) exhibit avoided crossings whereas levels of different symmetry exhibit genuine crossings.
	
	For odd integer $\kappa$ values, all eigenenergies corresponding to the periodic eigenstates $A$ are two-fold degenerate, see panel (b) of \cref{fig:zeta}. These degenerate states cannot be labeled by one of the specific symmetries, $A_1$ or $A_2$. Similarly, for even integer $\kappa$, the genuine crossings occur for the antiperiodic states $B_1$ and $B_2$, see panel (c) of \cref{fig:zeta}. In other words, if $\kappa$ is an odd (even) integer, the TISE (\ref{eq:RotTISE}) has two linearly independent solutions obeying the periodic (antiperiodic) boundary condition. This is referred to as coexistence of two linearly independent solutions with the same periodicity and is a peculiarity of the Whittaker-Hill equation (arising only for $\zeta>0$)  \cite{Djakov2005,Winkler1958,Magnus2004}. We note that the coexistence (degeneracy) of two Mathieu functions has been proved to be impossible \cite{Winkler1958}.   
	
	On the other hand, the avoided crossings occur between pairs of states with the symmetry $B_1$ or $B_2$ (i.e., between the energy curves with the same colors in panel (c) of \cref{fig:zeta}) for odd integer $\kappa$. For even integer $\kappa$, the avoided crossings occur between pairs of the $A_1$ or $A_2$ levels (i.e., between the energy curves with same colors in panel (b) of \cref{fig:zeta}). Note that some of the avoided crossings cannot be discerned on the scale of the figure. Therefore, one may conclude that the energy curves show extrema at even $\kappa$. Although this is valid for the lower energy levels, it is not always true for higher energy levels and larger $\kappa$ values.

	The discussion above regarding the energy levels of the generalized planar pendulum can be extended to the case of ultracold atoms in an optical superlattice, completing the picture of its semifinite-gap structure. In particular, the energy bands ($|\mathcal{D}(E)|\le2$) of \cref{eq:OPTISE} are intervals 
	\begin{align}
		[\epsilon^{(A)}_0 ,\epsilon^{(B)}_0 ], ~~ [\epsilon^{(B)}_1 ,\epsilon^{(A)}_1 ], ~~ [\epsilon^{(A)}_2 ,\epsilon^{(B)}_2 ], ~ \cdots 
	\end{align}
	separated by the gaps ($|\mathcal{D}(E)>2$) whose edges correspond to the GPP's eigenfunctions of the same periodicity: $A$-type gaps for periodic ($|q| = 0$) and $B$-type gaps for the antiperiodic ($|q| = 1/2$) boundary conditions, i.e., 
	\begin{align}
		(\epsilon^{(B)}_0 ,\epsilon^{(B)}_1 ), ~~ (\epsilon^{(A)}_1 ,\epsilon^{(A)}_2 ), ~~ (\epsilon^{(B)}_2 ,\epsilon^{(B)}_3 ), ~ \cdots 
	\end{align}
	Therefore, the genuine crossings in GPP's spectrum correspond to the closed gaps in the optical superlattice band structure. For even integer $\kappa$, all $B$-type gaps are closed except for the first $\kappa/2$. In addition, using the analytical energies (see \cref{sec:CQES},\ref{App1}), it is possible to derive analytic expressions for the widths of these $\kappa/2$ open $B$-type gaps. If $\kappa$ is an odd integer, all $A$-type gaps vanish except for the first $(\kappa-1)/2$, whose widths can be calculated analytically. The semifinite-gap structure for two examples, $\kappa=2$ and $\kappa=3$, are shown, respectively, in panels (a) and (b) of \cref{fig:HillD}. 
	\begin{figure}
		\centering
		\includegraphics[scale=0.45]{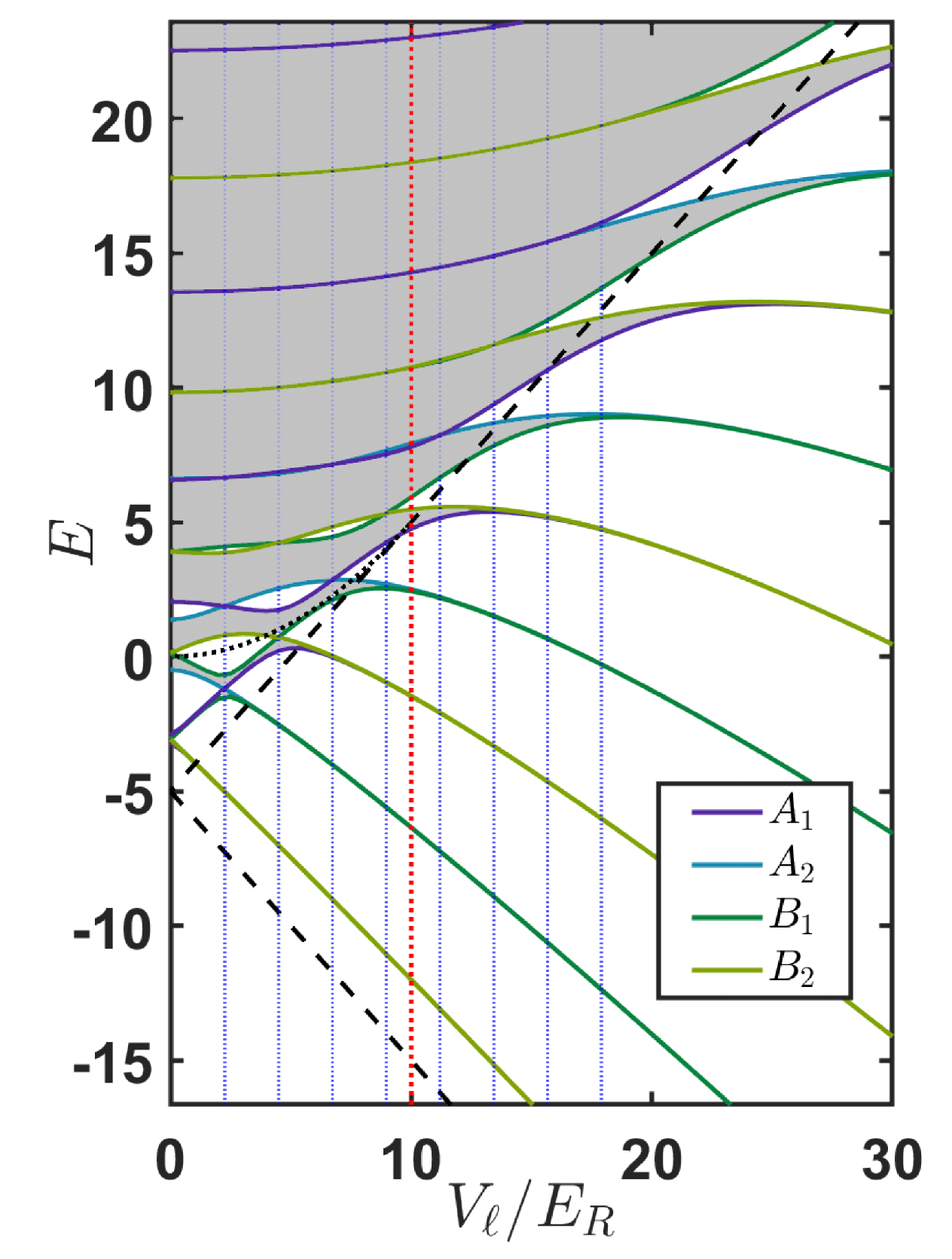} 
		\caption{\label{fig:BGzeta5}The band-gap structure of the optical superlattice 
			with respect to the long-lattice depth when the short-lattice depth is constant $V_s=-5~E_R$. The allowed bands are shaded in grey. Blue vertical dotted lines mark $V_\ell$ corresponding to $\kappa=1$ to $\kappa=8$. The red dotted line at $V_\ell = 2|V_s| = 10~E_R$ distinguishes the double-well (left side) and single-well (right side) regimes. For the single-well regime, black dashed lines indicate the maximum and minimum of the potential. For the double-well regime, see \cref{fig:zeta}. } 
	\end{figure} 
	
	The band structure of atoms in an optical superlattice for constant $V_s = -5~E_R$ and different $V_\ell$ is shown in \cref{fig:BGzeta5}. Note that while the energy bands below the maximum of the potential (i.e., below the upper black dashed line) are hardly discernible, those sufficiently above the potential's maximum exhibit a significant width. However, the gaps shrink with the energy of the band. These differences are more prominent when the optical superlattice has deeper wells, as can be seen by comparing  \cref{fig:zeta} with \cref{fig:BGzeta5}. Furthermore, guided by the color-coding assigned to different $\Gamma$ symmetries, we can see that with every transition from a genuine crossing (i.e., $V_\ell$ corresponding to closed gaps), the symmetry of the lower and upper band-edge states involved is interchanged ($A_1 \leftrightarrow A_2$ or $B_1 \leftrightarrow B_2$). We note that even though the gaps decrease in the high energy limit, the gaps become zero only at the loci of integer $\kappa$. Although further into the single-well regime (on the right from the red dotted vertical line in \cref{fig:BGzeta5}) the avoided and genuine crossings in principle still occur, the characteristic features of the double-well regime fade out, see \cref{fig:BGzeta5}.  
	
\begin{figure}
		\centering
		\includegraphics[scale=0.45]{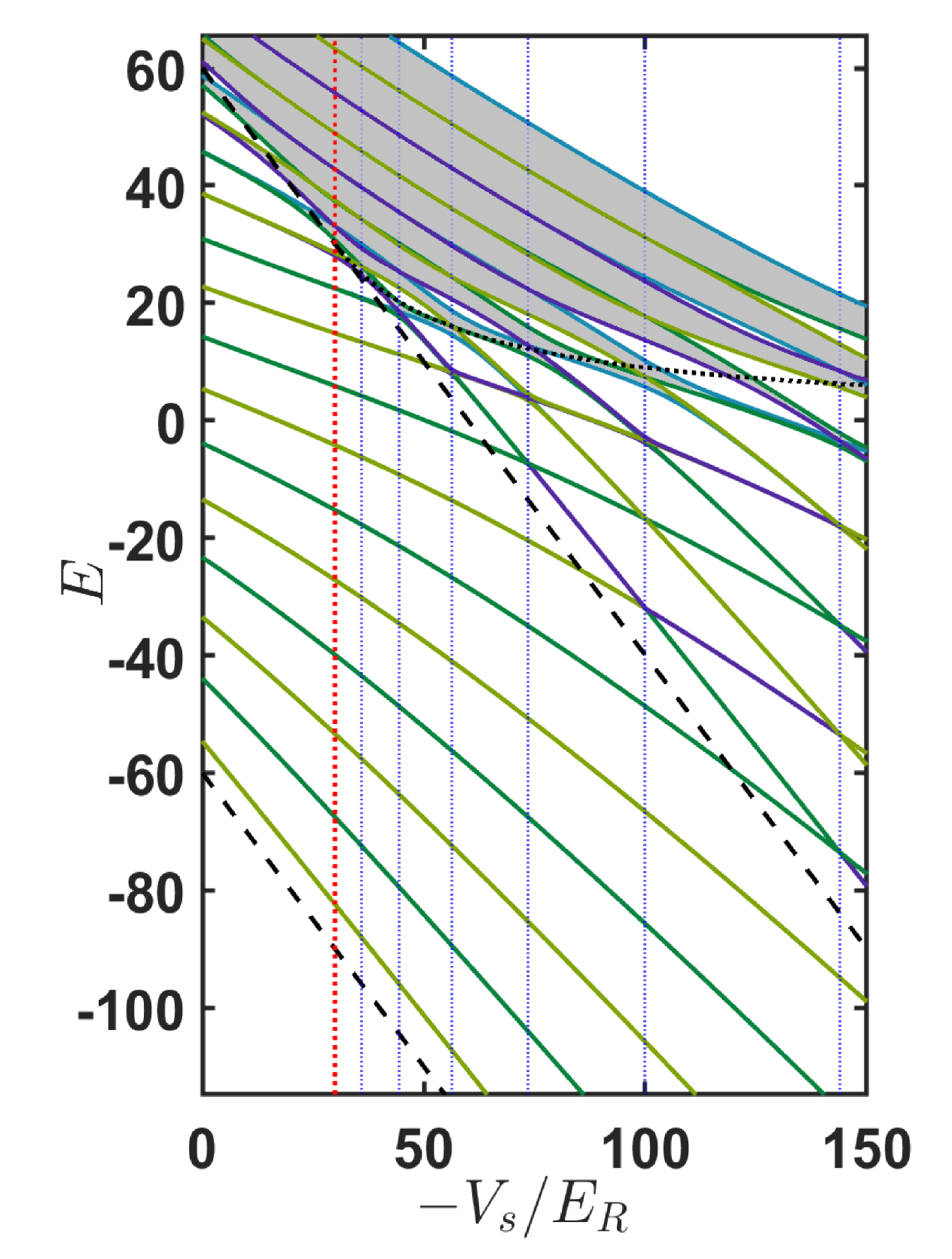} 
		\caption{\label{fig:BGeta60}The band-gap structure of atoms in an optical superlattice 
			with respect to the short-lattice depth when the long-lattice depth is constant $V_\ell=60~E_R$. Blue dotted lines mark $V_s$ corresponding to $\kappa$ from $\kappa=5$ to $\kappa=10$. The red dotted line at $-V_s = V_\ell/2 = 30~E_R$ separates the single-well (left side) and double-well (right side) regimes. In the single-well regime, black dashed lines indicate the maximum and minimum of the potential. For the double-well regime, the local and global minima are shown, respectively, by black dashed lines, and the maximum by the black dotted curve, cf. \cref{fig:zeta}. The color-coding is the same as in \cref{fig:BGzeta5}.} 
	\end{figure}

	\cref{fig:BGeta60} complements the overview of the above phenomena by displaying the band structure for a long-lattice well-depth $V_\ell=60~E_R$. The rich energy structure in the double-well regime ($|V_s|>30~E_R$, to the left of the red dotted vertical line) compared to the single-well regime ($|V_s|<30~E_R$) is clear in this figure where  the closed gaps located at $V_s = -144~E_R$, $V_s = -(60/7)^2\approx -73.47~E_R$, and $V_s = (60/9)^2\approx -44.44~E_R$ (i.e., $\kappa=5, 7 ,9$ are indicated by vertical dotted blue lines).       
		
	\section{Correspondence between orientation/alignment of a planar pendulum and spatial localization (squeezing) of an atom in an optical superlattice}\label{sec:Sloc}
	
	The concept of directionality (orientation and alignment) of a generalized planar pendulum corresponds to the spatial squeezing of atoms in an optical superlattice (see, e.g., Refs.~\cite{Leibscher2002,Leibscher2009}). In order to illustrate this correspondence, we make use of the common measures of orientation and alignment defined, respectively, as the expectation values $\langle\cos\theta\rangle$ and $\langle\cos^2\theta\rangle$. A fully oriented and fully anti-oriented planar pendulum is characterized, respectively, by $\langle\cos\theta \rangle = 1$ and $\langle\cos\theta \rangle=-1$. A fully aligned planar pendulum satisfies $\langle\cos^2\theta \rangle = 1$ whereas the spatial distribution of the axis of a free planar rotor (when the orienting and aligning interactions are absent) is characterized by the isotropic value $\langle\cos^2\theta \rangle = 1/2$.
	
\begin{figure}
		\centering
		\includegraphics[scale=0.3]{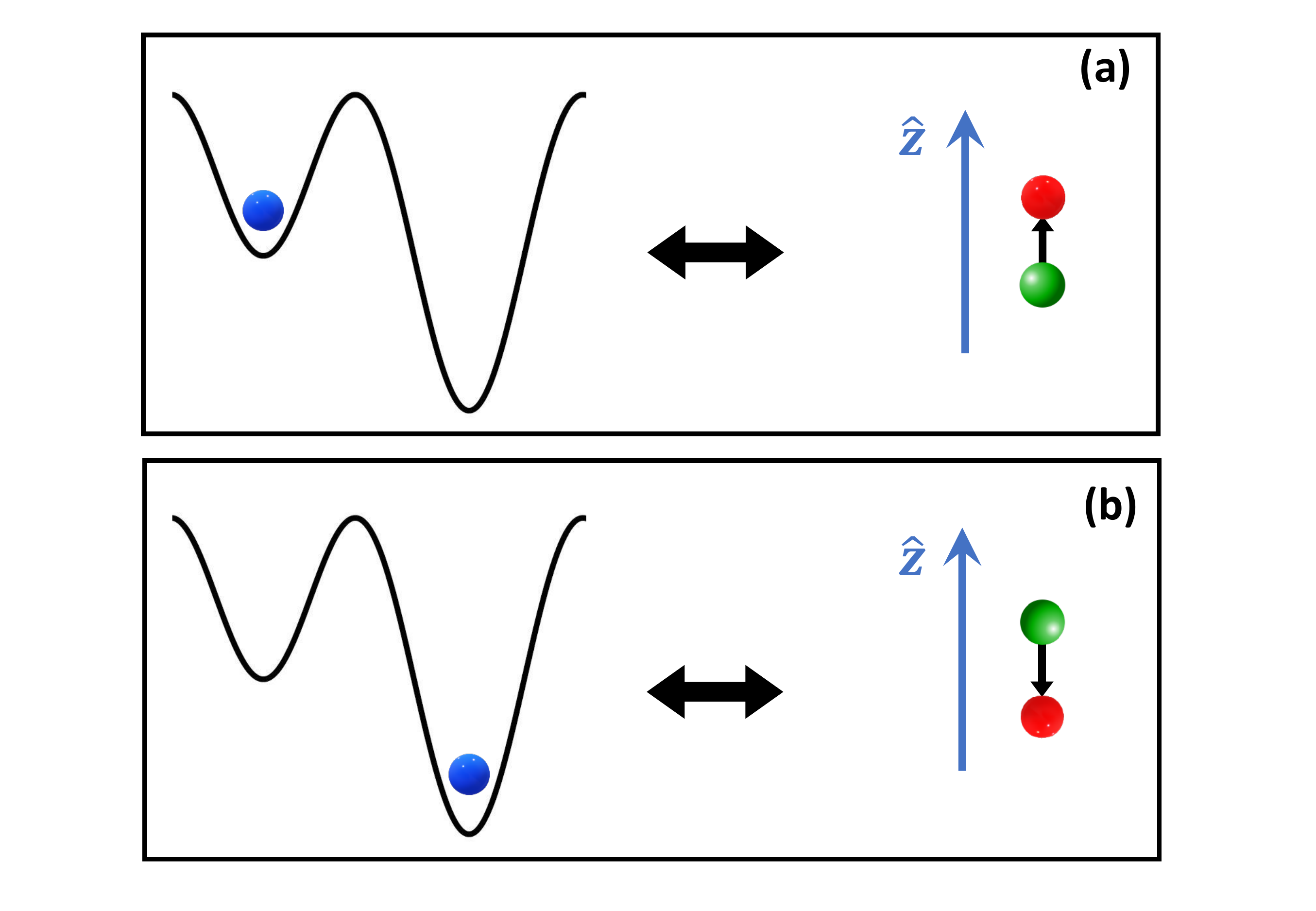} 
		\caption{\label{fig:OLvsRotor} Correspondence between orientation of a planar pendulum and spatial localization (squeezing) of an atom in an optical superlattice with $V_\ell>0$ and $V_s<0$. Panel (a) shows the oriented pendulum ($\theta = 0$) and panel (b) the anti-oriented pendulum ($\theta = \pi$). Here, $\hat{z}$ indicates the direction of the collinear external fields that turn a planar rotor into a generalized planar pendulum. } 
	\end{figure}

	Therefore, when the planar pendulum is oriented, $\theta \approx 0$, whereas when it is anti-oriented, $\theta \approx \pi$. Similarly, we find  alignment when $\theta \approx 0$ or $\pi$ and anti-alignment for $\theta \approx \pi/2$. \Cref{fig:OLvsRotor} shows a schematic representation of the relationship between orientation and alignment of the pendulum and the spatial squeezing of atoms in an optical superlattice with $V_\ell>0$ and $V_s<0$. As illustrated in the figure, an oriented planar pendulum is equivalent to the case of $\theta_a\approx 0$, i.e., the spatial localization of the atomic wavefunction (probability density) at the local minimum of the lattice. On the contrary, an anti-oriented planar pendulum is analogous to the case of spatial localization at the global minimum\footnote{Note that by choosing $V_\ell<0$, localization around the local/global minimum would be analogous to antiorientation/orientation.}. 
	
	\begin{figure}
		\centering
		\subfigure[$|\phi_0^{(A_1)}(\theta)|^2$ (ground state) for $\eta=-7~B$ equivalent to $V_\ell=7~E_R$  (i.e., $\kappa=1$).]{\includegraphics[scale=.45]{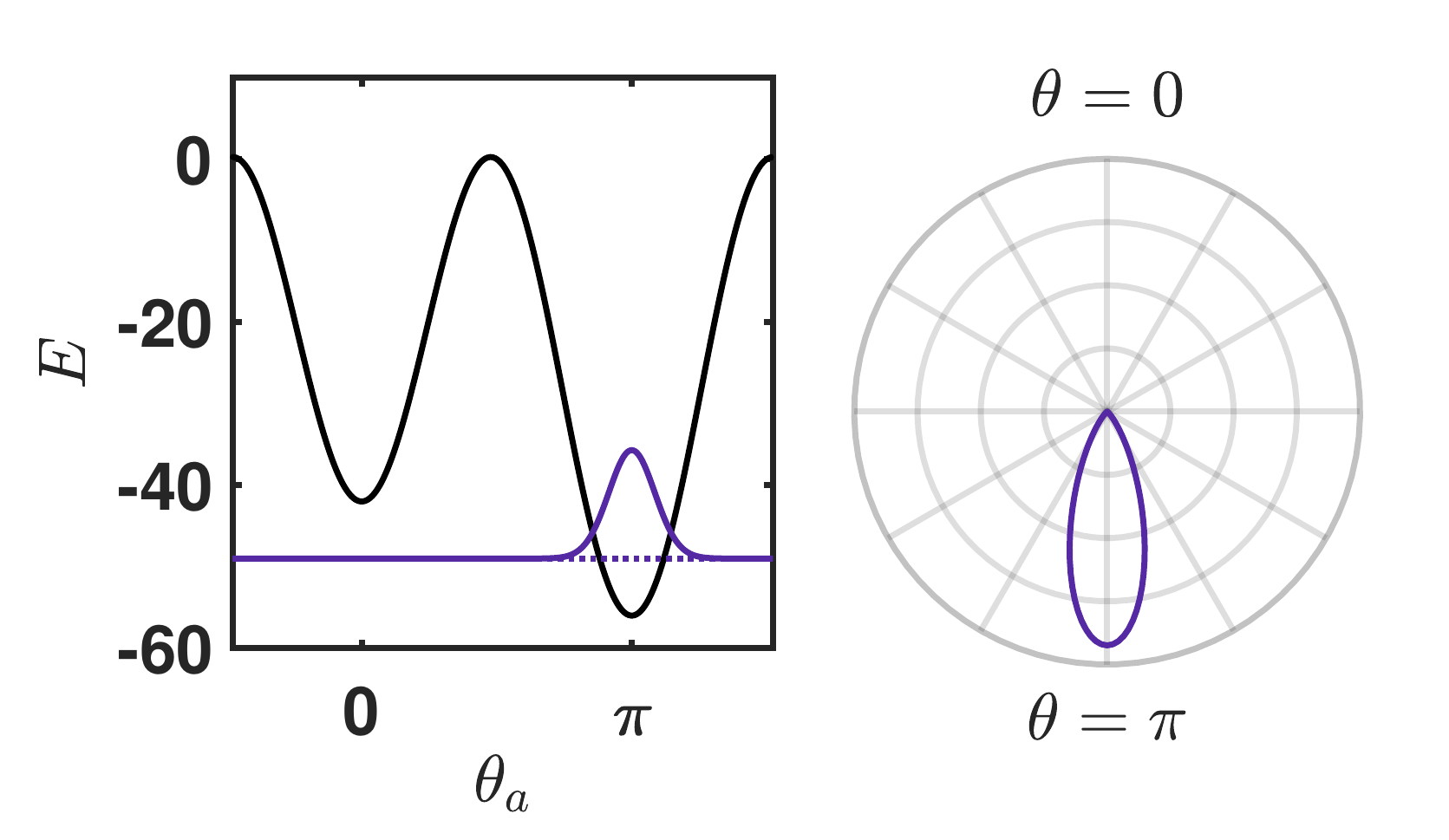}} 
		\subfigure[$|\phi_6^{(A_1)}(\theta)|^2$ for $\eta=-14~B$ equivalent to $V_\ell=14~E_R$ (i.e., $\kappa=2$). ]{\includegraphics[scale=.45]{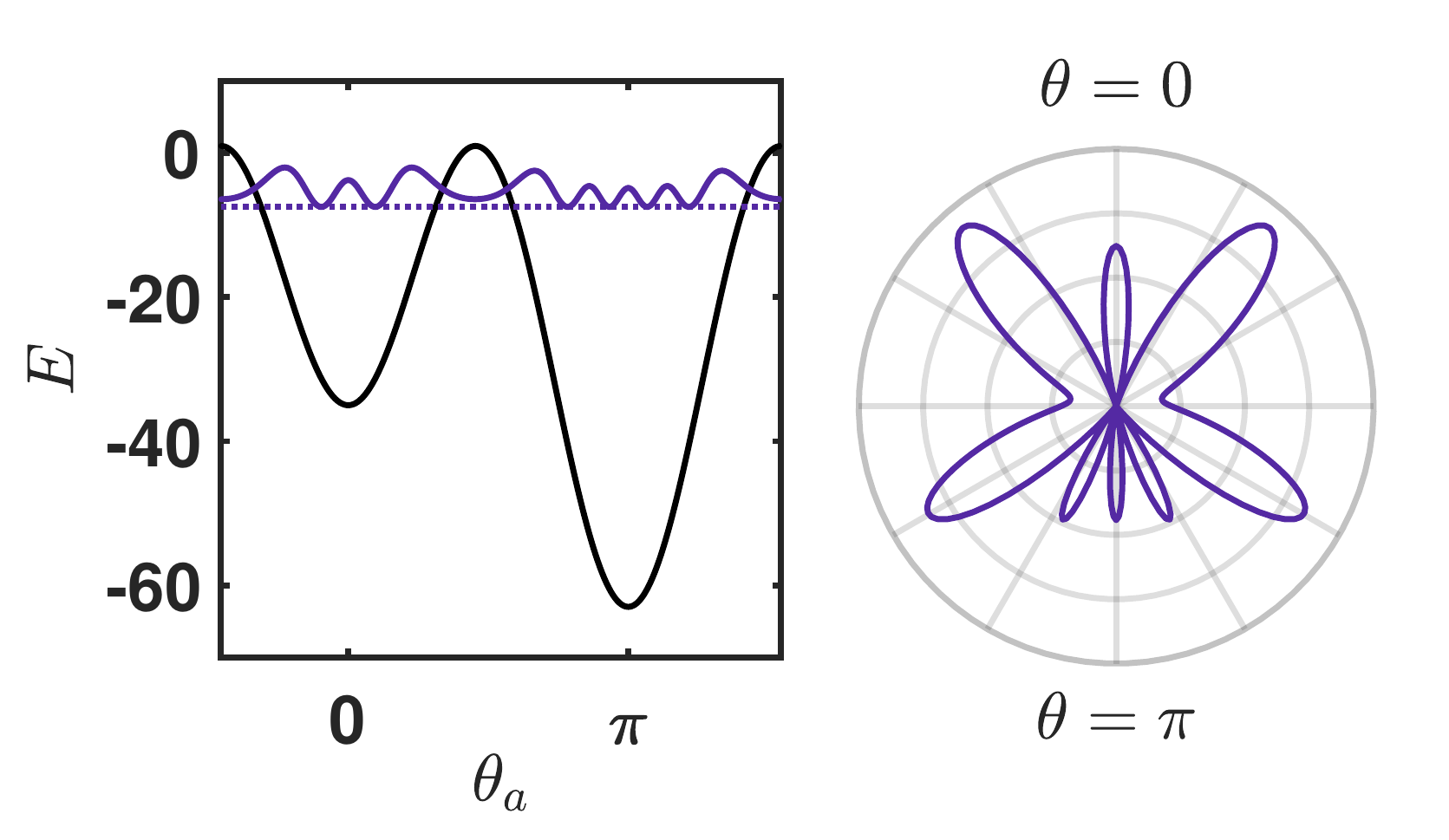}} 
		\caption{Plots of the probability density $|\psi(\theta_a)|^2$ in the unit cell of an optical superlattice for $V_s = -49~E_R$, together with the polar plots of its planar pendulum analogue, $|\phi(\theta)|^2$ with $\zeta = 49~B$.}
		\label{fig:densities}
	\end{figure}

   \cref{fig:densities} shows the probability densities $|\psi(\theta_a)|^2$ of two different bound states of an atom in an optical superlattice and its planar pendulum analog $|\phi(\theta)|^2$: (a) a highly localized state around the global minimum, and (b) a nearly delocalized state. Panel (a) corresponds to $\langle\cos\theta \rangle = -0.964$ and  $\langle\cos^2\theta \rangle  = 0.931$, which can be  charactrized as an anti-oriented and aligned pendular state. Panel (b), on the other hand, corresponds to $\langle\cos\theta \rangle = 0.095$ and  $\langle\cos^2\theta \rangle = 0.459$, a characteristic of an almost isotropic state ($\langle\cos\theta \rangle \approx 0$).
   
\begin{figure}
   	\centering
   	\includegraphics[scale=0.45]{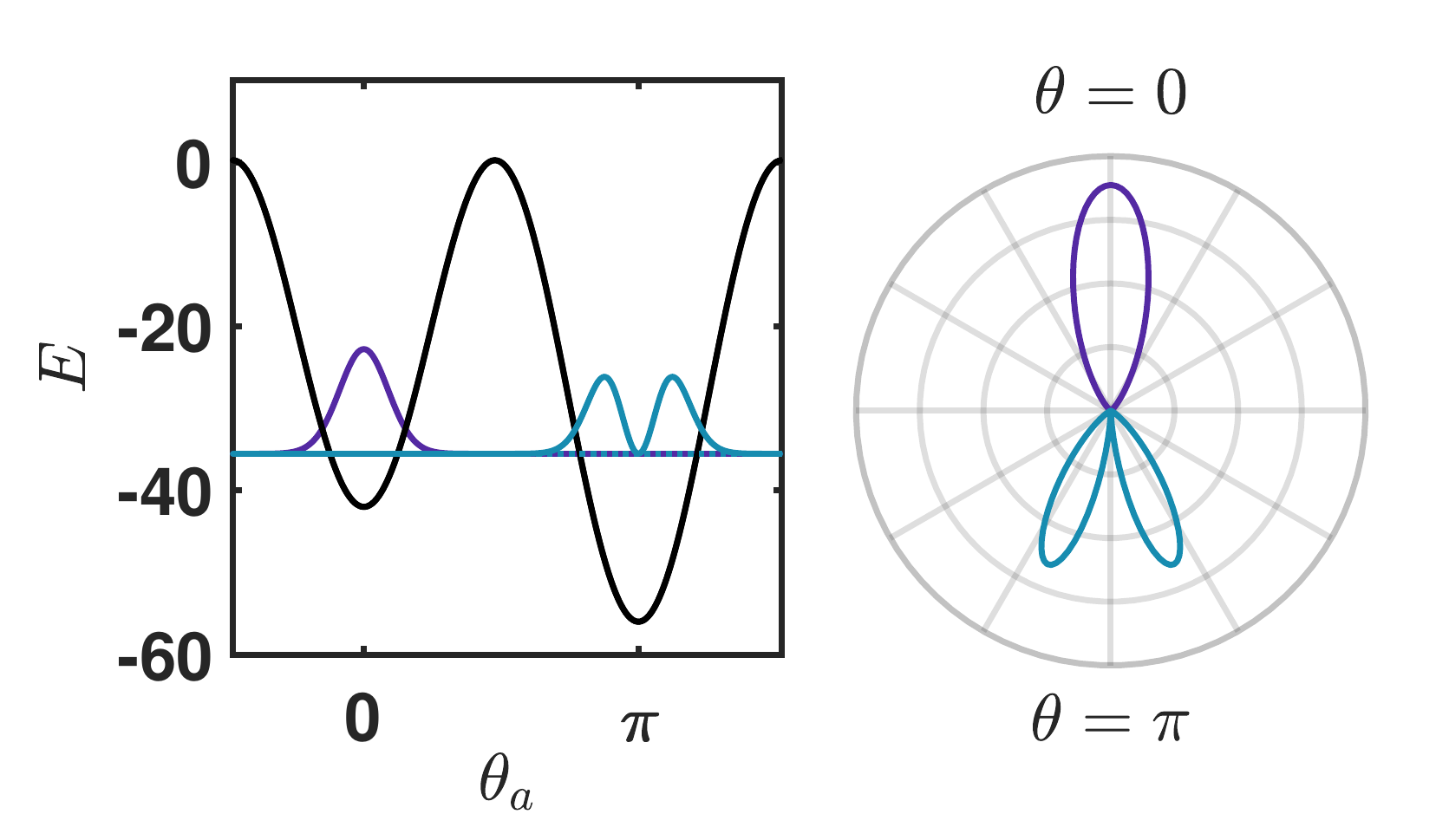} 
   	\caption{\label{fig:GC_dens_zeta49k1}Same as in \cref{fig:densities} but for a genuine crossing at $\zeta = 49~B$ and $\eta = -7~B$ or, equivalently, a closed gap at $V_s= -49~E_R, V_\ell = 7~E_R$. The color coding is the same as in \cref{fig:BGzeta5}.} 
   \end{figure}

    As mentioned before, the band-edge state at the closed gaps (i.e., a doubly degenerate state) results from a superposition of two driven rotor's states with different $\Gamma$ symmetries and hence, different localizations. An example is shown in \cref{fig:GC_dens_zeta49k1}, where the probability densities $|\phi_1^{(A_2)}(\theta)|^2$ and $|\phi_2^{(A_1)}(\theta)|^2$ corresponding to the lowest closed gap at $\kappa = 1$ are depicted. In this case, orientation and alignment of the state associated with $A_2$ symmetry (the light blue curve localized around the global minimum of the lattice) are $\langle\cos\theta \rangle  = -0.886$ and  $\langle\cos^2\theta \rangle = 0.793$, respectively. However, for the $A_1$ symmetry (the purple curve localized around local minimum of the lattice) we find $\langle\cos\theta \rangle  = 0.9603$ and  $\langle\cos^2\theta \rangle = 0.925$. Consequently, the superposition will be still (nearly) aligned but does have approximately zero orientation (double-arrow like). Note that the discussion given above also applies to the antiperiodic states. 
   
   Based on the correspondence between the GPP and OSL systems, we introduce the expectation values $\langle\cos\theta_a \rangle$ and  $\langle\cos^2\theta_a \rangle$ as quantitative measures of the spatial localization of the Bloch states in an optical superlattice. This makes it possible to use the Hellmann-Feynman theorem to establish a relationship between the spatial dependence of the band-edge energies, such as those shown in \cref{fig:zeta,fig:BGzeta5,fig:BGeta60}, and the spatial localization of the corresponding eigenstates \cite{Mirahmadi2020,Mirahmadi2021}. According to the Hellmann-Feynman theorem, $\left\langle \psi_{n} \left| \partial_\chi H(\chi) \right| \psi_{n} \right\rangle = \partial_\chi E_{n}$ for $\chi$ a parameter in the Hamiltonian, which in our case is either $V_\ell$ or $V_s$. Hence we obtain (over a single unit cell and for a constant $q$),
   \begin{align}\label{HF_eta}
   	\left\langle  \cos\theta_a  \right\rangle_n = \frac{\partial E_n}{\partial V_\ell} ~ \quad
   	\left\langle  \cos^2\theta_a  \right\rangle_n = \frac{\partial E_n}{\partial V_s}  
   \end{align}
   Therefore, the Hellmann-Feynman theorem implies that variations of the band energy in the vicinity of the genuine and avoided crossings will result in significant changes in the spatial localization of the atoms (see Ref.~\cite{Mirahmadi2020} for further details in the case of the pendulum). From \cref{HF_eta} and given that the global minimum of the potential $V(x)$ rises whereas the local minimum of the potential  drops at the avoided crossings (see \cref{sec:superlattice}), the localization of band-edge states around $\theta_a=0$ and $\theta_a=\pi$ interchanges, although the symmetry of the states involved remains the same.
   
   We note that the abrupt changes in the localization of the wavefunctions at these intersections are characteristic of energy levels well below the maximum of the potential but still above the local minimum. Indeed, for higher excited states, those changes occur more smoothly.
   
   \section{The role of the relative phase between the short and long lattice}\label{sec:delta_phi}
   
   We now examine the spectral properties of the OSL Hamiltonian (\ref{eq:Heff}) when the short and long lattice potentials are shifted by a non-zero phase $\varphi$, 
   \begin{equation}\label{eq:pot_phi}
   	V(\theta_a) = V_s \cos^2\theta_a + V_{\ell}\cos(\theta_a-\varphi)
   \end{equation}
   Since the Hamiltonian with potential (\ref{eq:pot_phi}) is not invariant under the parity transformation $\theta \mapsto -\theta$, its symmetry is not isomorphic with the $C_{2v}$ point group. In particular, the necessary condition for the degenerate states (genuine crossings) to occur, is not fulfilled. As a result, the semifinite-gap structure is not preserved for $\varphi\neq 0$. Moreover, in this case, the eigenproblem is no longer QES under the condition given in \cref{sec:CQES} and the topological index $\kappa$ should be redefined. The solvability of this problem is the subject of our forthcoming work.
   
\begin{figure}
   	\centering
   	\includegraphics[scale=0.55]{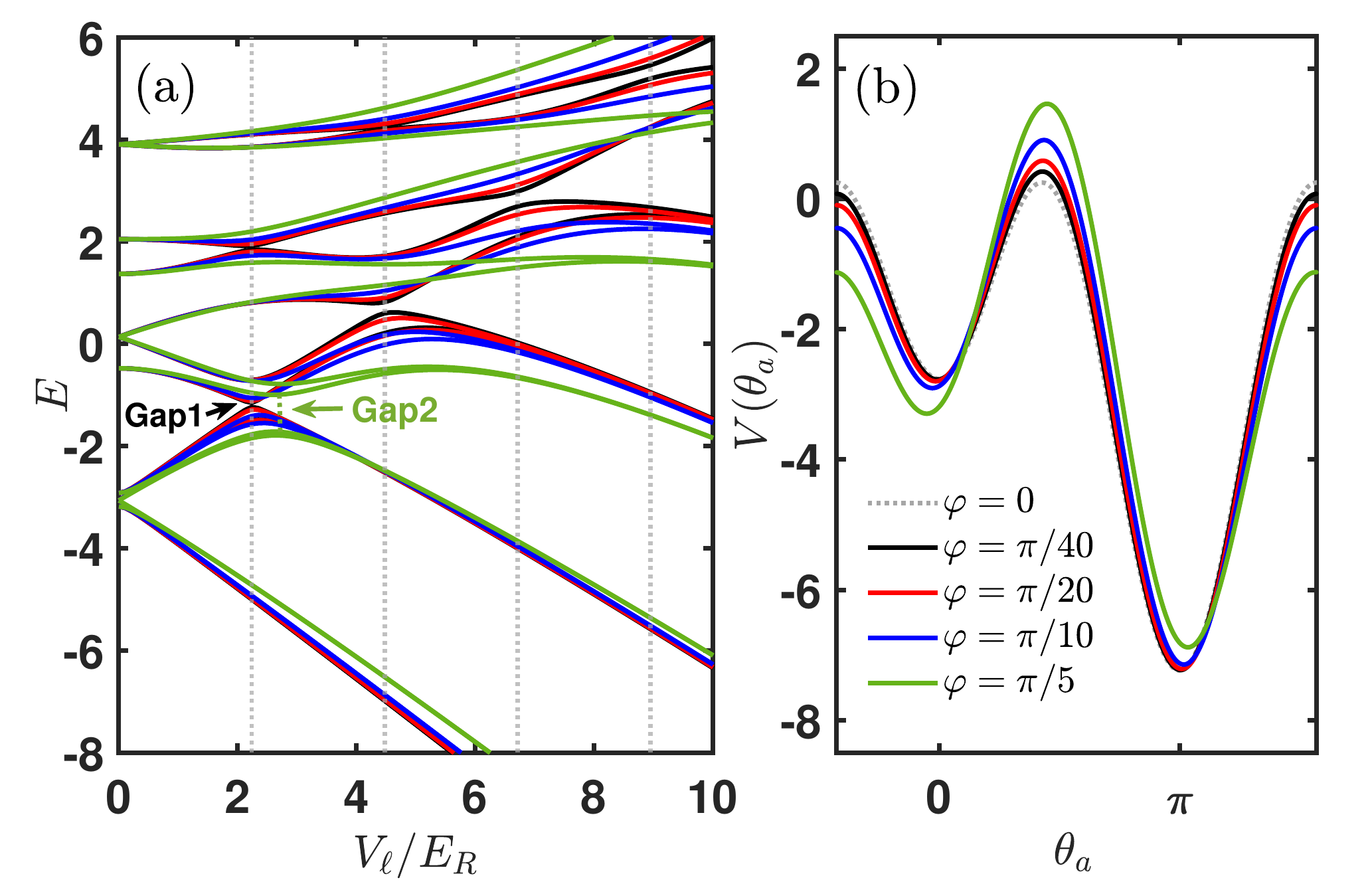} 
   	\caption{\label{fig:dphi}(a) Band-edge energy curves as a function of $V_\ell$ for the constant $V_s = -5~E_R$ for four different values of the relative phase $\varphi$. The gray dotted lines mark the $V_\ell$ values associated with integer ratios $\kappa=1 (V_\ell = \sqrt{5}~E_R)$ to 4 ($V_\ell = 4\sqrt{5}~E_R$). (b) Potential $V(\theta_a)$ for constant parameters $V_\ell=\sqrt{5}~E_R$ and $V_s = -5~E_R$ but different $\varphi$ values. } 
   \end{figure}

   However, using the WavePacket software, we calculated the band-edge energy curves for four different non-zero values of the phase $\varphi$. The results are displayed in \cref{fig:dphi}. Comparing panel~(a) with the double-well regime of \cref{fig:BGzeta5} (same potential but with $\varphi = 0$), we see that the band structure of the optical superlattice varies smoothly. By increasing $\varphi$, the closed gaps located at the loci of integer $\kappa$ (marked by gray dotted lines in panel~(a)) open and slowly shift to the right of the integer values (towards larger $\kappa$). Such changes occur over the whole band structure and become more pronounced at higher energies. 
   
   Although the OSL with $\varphi\neq 0$ do not have a semifinite-gap structure for the potential parameters corresponding to the integer $\kappa$, 
   for small values of $\varphi =\pi/40$ or $\varphi =\pi/20$ the gaps are still narrow and almost vanish for the smaller (integer) values of $\kappa$. An examples is marked as Gap1 in panel~(a) of \cref{fig:dphi} at $V_\ell=\sqrt{5}~E_R$ (i.e., $\kappa = 1$), showing the first gap in the spectrum of optical superlattices with relative phases $\varphi =\pi/40$ and $\varphi =\pi/20$. The corresponding potentials are plotted in the panel~(b) of \cref{fig:dphi}. It can be seen that the shape of these potentials (black and red curves) is almost preserved under the parity transformation $\theta_a \mapsto -\theta_a$,  resulting in a spectrum very similar to that of a lattice with $\varphi=0$ (dotted gray curve). By increasing $\varphi$ to $\pi/5$, the shape of the potential undergoes significant changes and the features apparent in the band structure of the for $\varphi = 0$ disappear. For instance, in this case, the narrowest (first) gap does not appear in the spectrum of the system with integer $\kappa$ (here $V_\ell=\sqrt{5}\approx 2.2$) but at $V_\ell\approx 2.7$ marked by Gap2 in panel~(a) of \cref{fig:dphi}.
   
   Also deeper optical superlattices have been examined with qualitatively the same results.
   
   We note that in previous work \cite{Folling2007,Calarco2004,Wang2013}, varying the relative phase $\varphi$ was used to control the lattice configuration.

	\section{Conclusions and prospects}\label{sec:concl}
	
	We have shown that two seemingly unrelated systems -- an atom under the potential of an optical superlattice and a planar rigid rotor under combined orienting and aligning interactions (also known as the generalized planar pendulum) -- have isomorphic Hamiltonians. We made use of this isomorphism and applied the extensive results obtained previously for the generalized pendulum based in part on the theory of the Whittaker-Hill equation \cite{Becker2017,Mirahmadi2020} to treat atoms in an optical superlattice.  Given that the generalized pendulum eigenproblem is conditionally quasi-exactly solvable, we have been able to obtain analytic results for atoms in an optical superlattice as well. 
	
	
	In particular, we have obtained in analytic form a finite number of eigenstates corresponding to the deep-lying band edges around the global minimum of the superlattice potential. Thereby, we prepared the soil for obtaining exact expressions for tunneling amplitudes between the sites of the superlattice (such as two global minima) and hence the hopping term in the corresponding Hubbard model Hamiltonian or the Landau-Zener tunneling probabilities \cite{Oberthaler2006}. By invoking the spectral properties of the Whittaker-Hill equation, we have shown that the motion of ultracold atoms in an optical superlattice gives rise to a semifinite-gap system that can be used to study topological properties of the atoms' energy spectra \cite{Lohse2016,Correa2008,Hemery2010}. Finally, we have shown how orientation and alignment of the generalized pendulum translate into the localization (squeezing) of the ultracold atoms in an optical superlattice. This treatment of atom squeezing offers itself to studying transport in optical superlattices \cite{Calarco2004}. 
	
	Conversely, the isomorphism between the generalized planar pendulum and the optical superlattice Hamiltonian would make it possible to simulate the planar rotor in the presence of external fields by the optical superlattice. In particular, ultracold atoms in an optical superlattice could be used to simulate the semifinite-gap spectrum of the supersymmetric partners of the planar rotor under the orienting and aligning interactions as well as under more involved potentials \cite{Lemeshko2011,Schmidt2014a,Schmidt2014}. Therefore, the present study can be viewed as a proposal for a quantum simulator of a planar rotor subject to external fields.
	
	In future work, the available analytic solutions will be used to develop analytic dynamical models of the trapping of atoms in an optical superlattice.
	
	We note that ultracold atoms in optical lattices are generally studied via the properties of the Mathieu equation that the time-independent Schr\"odinger equation for a simple 1D optical lattice $\propto \cos^2(kx)$ \cite{Windpassinger2013} reduces to. However, as we have shown herein, using the Whittaker-Hill equation instead, with its intriguing spectral features as well as its conditional quasi-exact solvability, reveals new perspectives on the optical superlattice eigenproblem that could prove useful in band structure engineering of ultracold quantum gases. This is a further indication that the fields of ultracold atoms, coherent control, and condensed matter physics are coming closer together.
	
	\section*{Acknowledgements}
		We thank Tommaso Macr\`i (QuEra Computing, Cambridge, MA), Mikhail Lemeshko (Institute of Science and Technology Austria), and Dominik Schneble (Stony Brook University) and his group for helpful discussions. We also thank Konrad Schatz (Society for the Advancement of Applied Computer Science, Berlin) for his insightful comments.  BF thanks John Doyle and Hossein Sadeghpour for their hospitality during his stay at Harvard Physics and at the Harvard \& Smithsonian Institute for Theoretical Atomic, Molecular, and Optical Physics.
	
	\appendix
	\section{Analytically obtainable band-edge states}\label{App1}
	The eigenstates of the GPP Hamiltonian can be obtained in analytic form by diagonalizing the four finite-dimensional symmetry-adapted matrix representations of this Hamiltonian. The solutions that satisfy the $q = 0$ (periodic boundary condition) can be written as
	
	\begin{align}\label{eig_A}
		\psi^{(A_1)}(\theta_a) =& \left(N^{(A_1)}\right)^{-1/2} e^{\beta\cos\theta_a}                    
		\sum_{\ell=0}^{(\kappa-1)/2}  v_{\ell} \cos^{2\ell}  \frac{\theta_a}{2} ~ \nonumber \\
		\psi^{(A_2)}(\theta_a) =&  \left(N^{(A_2)}\right)^{-1/2}
		e^{\beta\cos\theta_a} \sin\theta_a \sum_{\ell=0}^{(\kappa-3)/2} \tilde{v}_{\ell} \cos^{2\ell}\frac{\theta_a}{2} 
	\end{align}
	which are normalized by (on the $2\pi$ interval of $\theta_a$)
	    \begin{align}\label{N_A1A2}
			N^{(A_1)} &= 2\pi\sum_{\ell,\ell'} \frac{1 }{2^{2L}} v_{\ell} v_{\ell'} \left\lbrace \binom{2L}{L} I_0(2\beta) + 2\sum_{j=0}^{L-1} \binom{2L}{j} I_{L-j}(2\beta) \right\rbrace  ~ \nonumber \\
			N^{(A_2)} &= 2\pi\sum_{\ell,\ell'} \frac{1 }{2^{2L+1}} \tilde{v}_{\ell} \tilde{v}_{\ell'} \Bigg\{   \binom{2L}{L} I_1(2\beta)/\beta + \sum_{j=0}^{L-1} \binom{2L}{j} [ 2I_{L-j}(2\beta)  - I_{L-j+2}(2\beta) \nonumber \\
			&- I_{L-j-2}(2\beta)]  \Bigg\} 
		\end{align} 
	The constants $v_{\ell}$ and $\tilde{v}_{\ell}$ are components of the eigenvectors of the matrix representations corresponding to the $A_1$ or $A_2$ symmetries (see Refs.~\cite{Becker2017,Mirahmadi2020}). $I_{\rho}$ is the modified Bessel function of the first kind and $\rho$th order \cite{Arfken2005,Gradshteyn2007}, $\binom{b}{a}$ is the binomial coefficient, and $L\coloneqq\ell+\ell'$. 
	
\begin{figure}
		\centering
		\includegraphics[scale=0.4]{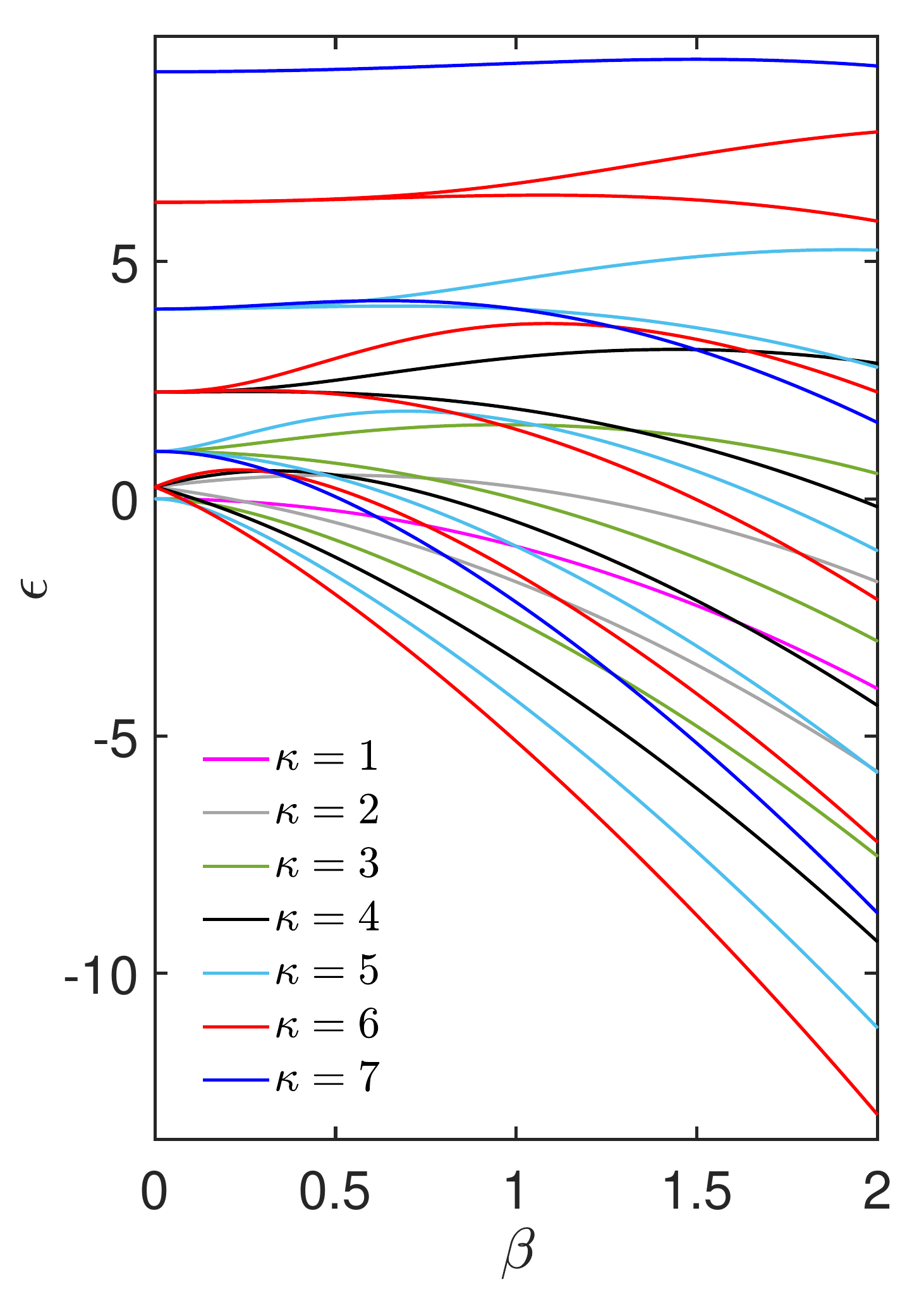} 
		\caption{\label{fig:eigen_beta} The lowest available analytic eigenvalues of the generalized planar pendulum (band-edge energies in optical superlattice) as functions of $\beta = \sqrt{\zeta/B} = \sqrt{-V_s/E_R}$ for different integer values of the topological index $\kappa$ given by \cref{eq:kappa}.} 
	\end{figure}

	The $2\pi$-antiperiodic solutions can be written as 
	\begin{align}\label{eig_B}
		\psi^{(B_1)}(\theta_a) & = \left(N^{(B_1)}\right)^{-1/2} e^{\beta\cos\theta_a} \cos\frac{\theta_a}{2}                    
		\sum_{\ell=0}^{(\kappa-2)/2}  w_{\ell} \cos^{2\ell}  \frac{\theta_a}{2} ~, \nonumber \\
		\psi^{(B_2)}(\theta_a) & =  \left(N^{(B_2)}\right)^{-1/2} 
		e^{\beta\cos\theta_a} \sin\frac{\theta_a}{2} \sum_{\ell=0}^{(\kappa-2)/2} \tilde{w}_{\ell} \cos^{2\ell}\frac{\theta_a}{2}  ~
	\end{align}
	where, the constants $w_{\ell}$ and $\tilde{w}_{\ell}$ are components of the eigenvectors of the matrix representations associated with the $B_1$ and $B_2$ symmetries given in Refs.~\cite{Becker2017,Mirahmadi2020}.
	The normalization constants are 
		\begin{align}\label{N_B1B2}
			N^{(B_1)} &= 2\pi\sum_{\ell,\ell'} \frac{1 }{2^{2L+1}} w_{\ell} w_{\ell'} \Bigg\lbrace  \binom{2L}{L} \left[ I_0(2\beta) + I_1(2\beta) \right]  + \sum_{j=0}^{L-1} \binom{2L}{j} [ 2I_{L-j}(2\beta) \nonumber \\
			&+ I_{L-j+1}(2\beta) + I_{L-j-1}(2\beta) ] \Bigg\rbrace  ~ \nonumber \\
			N^{(B_2)} &= 2\pi\sum_{\ell,\ell'} \frac{1 }{2^{2L+1}} \tilde{w}_{\ell} \tilde{w}_{\ell'} \Bigg\lbrace  \binom{2L}{L} \left[ I_0(2\beta) - I_1(2\beta) \right]  + \sum_{j=0}^{L-1} \binom{2L}{j} [ 2I_{L-j}(2\beta) \nonumber \\
			&- I_{L-j+1}(2\beta) - I_{L-j-1}(2\beta) ] \Bigg\rbrace ~
		\end{align}
	A total of 40 analytic solutions are given in  Refs.~\cite{Becker2017,Mirahmadi2020}. \Cref{fig:eigen_beta} displays 24 of the analytic energy curves as a functions of $\beta$ for different values of $\kappa$. It is important to keep in mind that the superlattice geometry changes from a single-well for $\beta<\kappa/2$ to a double-well per site for $\beta>\kappa/2$. 

	\section{The Hill discriminant}\label{App2}
	Consider the differential equation 
	\begin{align}\label{eq:appB_0}
		\frac{d^2f}{dy^2} + [\lambda+Q(y)]f(y) = 0~
	\end{align}
	where $Q(y+T) = Q(y)$ is a real-valued smooth periodic function and $\lambda$ is the eigenvalue. This periodic differential equation (often called the Hill equation~\cite{Magnus2004}) has a band-gap structure. Due to the translational symmetry, functions $f(y)$ should fulfil the following boundary condition
	\begin{align}\label{eq:appB_1}
		f(y+T) = \mu f(y)~
	\end{align}
	where $\mu$ is known as the Floquet multiplier. 
	
	In order to study its spectrum, we choose a basis set consisting of two linearly independent solutions $f_1(y,\lambda)$ and $f_2(y,\lambda)$ corresponding to the same eigenvalue $\lambda$ and obeying the conditions $f_1(0,\lambda) = f_2'(0,\lambda) = 1$ and $ f_2(0,\lambda) = f_1'(0,\lambda) = 0$ (prime denotes the derivative with respect to $y$). Defining the general $f$ function corresponding to the eigenvalue $\lambda$ as
	
	\begin{align}\label{eq:appB_2}
		f(y,\lambda) = \alpha f_1(y,\lambda) + \beta f_2(y,\lambda)
	\end{align}
	and substituting it into \cref{eq:appB_1} and its derivative, we have
	\begin{align}\label{eq:appB_3}
		\begin{bmatrix}
			f_1(T,\lambda) & f_2(T,\lambda)\\
			f_1'(T,\lambda) & f_2'(T,\lambda)
		\end{bmatrix}
		\begin{bmatrix}
			\alpha \\
			\beta
		\end{bmatrix}
		= \mu
		\begin{bmatrix}
			\alpha \\
			\beta
		\end{bmatrix}~
	\end{align}
	The 2-by-2 matrix on the left hand side of \cref{eq:appB_3} is the transpose of the monodromy matrix with a constant ($y$-independent) determinant equal to one~\cite{Teschl2012g,Correa2008,Magnus2004}. It is known that the eigenvalues of the monodromy matrix, $\mu$, are Floquet multipliers and that the trace of the monodromy matrix,
	\begin{align}\label{eq:appB_4}
		\mathcal{D}(\lambda)=f_1(T,\lambda)+f_2'(T,\lambda)~,
	\end{align}
	is the discriminant associated with the Hill equation~\eqref{eq:appB_0}, the so-called Hill discriminant or Floquet discriminant~\cite{Correa2008,Hemery2010}. Note that some authors define $\mathcal{D}(\lambda)$ as half of this value (see e.g. Refs.~\cite{Kohn1959,Teschl2012g}). Thus, from \cref{eq:appB_3}, the characteristic equation for the eigenvalues $\mu$ reduces to
	\begin{align}\label{eq:appB_5}
		\mu^2-\mathcal{D}(\lambda)\mu+1=0 ~
	\end{align}
	
	By substituting $\mu=\exp(i T q)$ in \cref{eq:appB_5}, we obtain the relation between the Hill discriminant and the $q$ parameter as
	\begin{align}\label{eq:appB_6}
		\mathcal{D}(\lambda) = 2\cos(T q(\lambda))~
	\end{align} 
	Finally, the spectral properties of \cref{eq:appB_0} can be explained by using the oscillation properties of its associated Hill discriminant $\mathcal{D}(\lambda)$~\cite{Kramers1935,Kohn1959,Magnus2004,Correa2008}:  (i) for $|\mathcal{D}(\lambda)|\le2$, the parameter $q$ is real and so the modulus of $\mu$ is equal to one. Therefore, the solutions $f(y)$ are bounded and the corresponding $\lambda$ values are allowed (bands).     
	(ii) for $|\mathcal{D}(\lambda)|>2$, the parameter $q$ is not real and so the solutions $f(y)$ do not have a finite norm and thus are not physically admissible. The corresponding $\lambda$ values are forbidden (gaps).    
	Furthermore, the $\lambda$ values with $\mathcal{D}(\lambda)=2$ are the spectral edges corresponding to the periodic solutions ($\mu = 1$), and those with $\mathcal{D}(\lambda)=-2$ to the antiperiodic solutions ($\mu=-1$). Therefore, these values describe the edges of the allowed regions (band-edges).
	
	We note  that the oscillations of the discriminant $\mathcal{D}(\lambda)$ as a function of real eigenvalues $\lambda$ and their intersections with lines
	$\mathcal{D}(\lambda)=\pm2$ depend on the shape of the periodic function $Q(y)$ in \cref{eq:appB_0}.

	\bibliographystyle{iopart-num-href}
	\bibliography{OSL.bib}

\end{document}